\documentclass[landscape,usenatbib,usegraphicx]{mn2e}
\usepackage{graphicx}
\usepackage{colordvi}

\newcommand{\tk}{\tilde{k}}
\def\ba{\begin{eqnarray}}
\def\ea{\end{eqnarray}}
\newcommand{\be}{\begin{eqnarray}}
\newcommand{\ee}{\end{eqnarray}}

\title[The contribution of PMFs to CMB anisotropies]{The scalar, vector and tensor contributions of a stochastic background of magnetic fields to CMB anisotropies}

\author[D. Paoletti, F.Finelli,F.Paci]
{D.~Paoletti \thanks{paoletti@iasfbo.inaf.it}$^{1,2}$, 
F.~Finelli \thanks{finelli@iasfbo.inaf.it}$^{2,3,4}$ and 
F.~Paci \thanks{paci@iasfbo.inaf.it}$^{5,2,4}$
\\
$^1$ Dip. di Fisica, Universit\`a degli studi di Ferrara and INFN, \\
via Saragat 1, I-44100 Ferrara - Italy \\
$^2$ INAF-IASF Bologna, Istituto di Astrofisica Spaziale e Fisica Cosmica 
di Bologna \\
Istituto Nazionale di Astrofisica, via Gobetti 101, I-40129 Bologna - Italy \\
$^3$ INAF-OAB, Osservatorio Astronomico di Bologna \\
Istituto Nazionale di Astrofisica,
via Ranzani 1, I-40127 Bologna - Italy\\
$^4$ INFN, Sezione di Bologna,
Via Irnerio 46, I-40126 Bologna, Italy \\
$^5$ Dipartimento di Astronomia, Universit\`a degli Studi di Bologna,\\
via Ranzani, 1 -- I-40127 Bologna -- Italy}

\begin{document}

\def\lsim{\,\lower2truept\hbox{${< \atop\hbox{\raise4truept\hbox{$\sim$}}}$}\,}
\def\gsim{\,\lower2truept\hbox{${> \atop\hbox{\raise4truept\hbox{$\sim$}}}$}\,}

\maketitle

\begin{abstract}
We study the contribution of a stochastic background (SB) of primordial 
magnetic fields (PMF) on the anisotropies in temperature 
and polarization of the 
cosmic microwave background radiation (CMB). A SB of PMF 
modelled as a fully inhomogeneous component 
induces non-gaussian scalar, vector and tensor metric linear perturbations. 
We give the exact expressions for the Fourier spectra 
of the relevant energy-momentum components of such SB, 
given a power-law dependence parametrized by a spectral index $n_B$ 
for the magnetic field power spectrum cut at a damping scale $k_D$. 
For all the values of $n_B$ considered here, the contribution to the 
CMB temperature pattern by such a SB is dominated by the 
scalar contribution and then by the vector one at higher multipoles. 
We also give an analytic estimate of the scalar contribution to the 
CMB temperature pattern.
\end{abstract}


\raggedbottom
\setcounter{page}{1}
\section{Introduction}
\setcounter{equation}{0}
\label{intro}
The origin of the large scale magnetic fields observed is an issue of 
great importance in astrophysics (see \cite{subra_review} for a review). 
Primordial magnetic fields (PMF) generated in the early Universe 
are a possible explanation of large scale magnetic fields in clusters of galaxies and galaxies
and might have left an imprint in the 
anisotropy pattern of the cosmic microwave background (CMB) (see 
\cite{durrer_review} for a review). 

A stochastic background (SB) is the simplest way to model 
these random PMFs in an isotropic and homogeneous background.
A SB of PMF is modelled as a fully inhomogeneous component 
and its energy momentum tensor (EMT) - quadratic in the magnetic fields -
is considered at the same footing as linear inhomogeneities in the 
other components and linear metric fluctuations. 
A SB of PMF generates independent modes for 
all kinds of linear perturbations: there has been several studies 
for scalar 
(\cite{KL,KR,giovanninikunze} and references therein; 
Finelli, Paci \& Paoletti 2008 - henceforth \cite{FPP_PRD} -),  
vector \citep{SB,SS,MKK,lewis} and tensor \citep{DFK,MKK,CDK} perturbations in 
presence of a SB of PMF. Limits on the amplitude and spectral index of 
the PMF were also obtained by an exploration of a flat $\Lambda$CDM  model 
in presence of such SB \citep{yamazaki}.

%All kinds of perturbations generated by a SB of PMF 
%have been studied by several authors. 

We study the problem 
by a numerical Eistein-Boltzmann code extending the results 
obtained in our previous work \citep{FPP_PRD}.
The study of the impact of a SB of PMF on CMB anisotropies requires 
a detailed study of the initial conditions for fluctuations and of the 
power spectra of the EMT of the SB of PMF. 
Our paper improves previous results in these two aspects.
In this paper 
we obtain the Fourier spectra of the relevant
vector and tensor energy-momentum components of the SB of PMF along the
procedure used in \cite{FPP_PRD} for the scalar components. As shown in 
\cite{FPP_PRD}, by solving exactly the convolution integrals for a 
sequence of values for the spectral slope $n_B$ which parametrizes the 
PMF power spectrum, previous results of \cite{MKK}
may be significantly improved.

With these improved correlators we then investigate the impact 
of a stochastic background of primordial magnetic fields on 
scalar, vector and tensor cosmological perturbations and in particular on 
CMB temperature and polarization anisotropies. Our results show that it is very 
important to study also vector perturbations  
since these dominate at high $\ell$ over the scalar ones for any slope of the 
spectrum of PMF.

Our paper is organized as follows. In Section 2 we introduce our conventions 
for a non-helical SB and the scalar, vector and tensor decomposition of its 
EMT: we give our exact results for the relevant objects for a set 
of $n_B$ leaving the details in Appendices A, B, and we show how our 
exact results improve on previous results. The set of $n_B$ now 
includes also values $n_B < -3/2$ which were not studied in \cite{FPP_PRD}. 
Section 3 presents the decomposition of metric perturbations and 
sections 4,5,6 present the study of scalar, vector and tensor cosmological 
perturbations in presence of a SB of PMF, respectively. 
In Section 7 we present the results obtained by our modified version of 
the CAMB code \citep{CAMB}. We conclude in Section 8.

\section{Stochastic Background of Primordial Magnetic Fields}

Since the EMT of PMF at homogeneous level is zero,
at linear order PMFs evolve like a stiff source and therefore it
is possible to discard the back-reaction of gravity
onto the SB of PMF.
Before the decoupling epoch the electric conductivity of the
primordial plasma is very large,
therefore it is possible at lowest order to consider the
infinite conductivity limit, in which the induced electric field is zero.
Within the infinite conductivity limit the magnetic field amplitude scales 
simply as ${\bf B}({\bf x},\tau)={\bf B}({\bf x})/a(\tau)^2$ and
\footnote{We choose the standard convention in which at present time $t_0$,
$a(t_0) = 1$.}
%In the infinite conductivity limit (i.e. vanishing electric field) 
the EMT of a SB of PMF is:
\ba
\tau^0_0&=&-\rho_B=-\frac{B^2({\bf x})}{8\pi a^4(\tau)}\\
\tau_i^0&=&0\\
\tau_j^i&=&\frac{1}{4\pi a^4(\tau)}
\Big(\frac{B^2({\bf x})}{2}\delta_j^i-B_j({\bf x})B^i({\bf x})\Big)
\ea
The two point correlation function in the Fourier space 
\footnote{As Fourier transform and its inverse, we use:
\ba 
Y ({\bf k}, \tau)& = & \int d{\bf x}
e^{i {\bf k} \cdot {\bf x}} Y ({\bf x}, \tau) \nonumber \\
%\, ,\,\,\,\,\, 
Y ({\bf x}, \tau) &=& \int \frac{d{\bf k}}{(2 \pi)^3}
e^{-i {\bf k} \cdot {\bf x}} Y ({\bf k}, \tau) \,. \nonumber 
\label{Fourier}
\ea
where $Y$ is a generic function. Note that we have changed our Fourier 
conventions with respect to \cite{FPP_PRD}.}
for fully inhomogeneous fields is:
\be
\langle B_i({\bf k}) B_j^*({\bf k}')\rangle=(2\pi)^3 \delta({\bf k}-{\bf k}')
(\delta_{ij}-\hat k_i\hat k_j) \frac{P_B(k)}{2}
\ee
Where $P_B(k)$ is the spectrum of PMF parametrized as:
\be
P_B(k)=A \Big(\frac{k}{k_*}\Big)^{n_B}\,,
\label{PSpectrum}
\ee 
and $k_*$ is a reference scale.
PMF are damped on small scales by radiation viscosity. 
We model this damping 
introducing a sharp cut-off in the PMF power spectrum 
at a damping scale called $k_D$.
The relation between the amplitude of 
PMF power spectrum and the amplitude of the field itself is:
{\setlength\arraycolsep{2pt}
\ba
\langle B^2\rangle&=&\langle B_i^*({\bf x}') B_i({\bf x})\rangle|_{{\bf x}'={\bf x}}\nonumber\\
%&=& \int d {\bf k} \int d {\bf k}' e^{i ({\bf k}-{\bf k}'){\bf x}} \langle B_i^*({\bf k}') B_i({\bf k})\rangle\nonumber\\
%&=& \int d {\bf k} \int d {\bf k}' e^{i ({\bf k}-{\bf k}'){\bf x}} \delta({\bf k}-{\bf k}') (\delta_{ii}-k_ik_i) \frac{P_B(k)}{2}\nonumber\\
&=&  \frac{1}{2 \pi^2} \int_0^{k_D} dk\, k^2 P_B(k)
\ea}
Solving the integral above we obtain:
\be
\langle B^2\rangle= \frac{A}{2 \pi^2 (n_B+3)} \frac{k_D^{n_B+3}}{k_*^{n_B}} \,,
\ee
where for the convergence of the integral above is requested $n_B>-3$. 
We shall use to denote the
magnetic field amplitude this quantity instead of smearing the field at an 
additional scale $\lambda$,
as in \cite{FPP_PRD}:
\be
\langle B^2\rangle_\lambda=\frac{1}{2\pi}\int dk k^2 P_B(k)e^{-\lambda^2 k^2}
\ee 
The two definitions can however related simply by:
\be
\langle B^2\rangle=\langle B^2\rangle_\lambda \frac{k_D^{n_B+3}\lambda^{n_B+3}}{(n_B+3)\Gamma\Big(\frac{n_B+3}{2}\Big)}
\ee

The EMT of PMF is quadratic in the magnetic field and therefore its Fourier 
transfom is a convolution. 
The two point correlation function of the spatial part of 
EMT is\footnote{We use the convention that latin indexes run 
from 1 to 3 while greek indexes run from 0 to 3}:
%{\setlength\arraycolsep{2pt}
\ba
\langle\tau^*_{ab}({\bf k})\tau_{cd}({\bf k'})\rangle&=&
%\frac{1}{64\pi^5}
\int \frac{ d {\bf q} d{\bf p}}{64\pi^5} \delta_{ab}\delta_{cd}\langle B_l({\bf q})
B_l({\bf k}-{\bf q})B_m({\bf p})B_m({\bf k}'-{\bf p})\rangle\nonumber\\
&& - \int \frac{ d {\bf q} d{\bf p}}{32\pi^5}
\langle B_a({\bf q})B_b({\bf k}-{\bf q})B_c({\bf p})
B_d({\bf k}'-{\bf p})\rangle\nonumber
\ea
%}
We can then obtain scalar, vector and tensor correlation functions:
\ba
&&\langle\Pi^{*(S)}({\bf k})\Pi^{(S)}({\bf k'})\rangle 
= \delta_{ab}\delta_{cd}\langle\tau^*_{ab}({\bf k})\tau_{cd}({\bf k'})\rangle\nonumber\\
&&\langle\Pi_{i}^{*(V)}({\bf k})\Pi_j^{(V)}({\bf k'})\rangle 
= k_a P_{ib}({\bf k}) k'_c P_{jd}({\bf k'})\langle\tau^*_{ab}({\bf k})
\tau_{cd}({\bf k'})\rangle\nonumber\\
&&\langle\Pi_{ij}^{*(T)}({\bf k})\Pi_{tl}^{(T)}({\bf k'})\rangle 
= (P_{ia}({\bf k})P_{jb}({\bf k})-\frac{1}{2} P_{ij}({\bf k})P_{ab}({\bf k}))
\times\nonumber\\
&&(P_{tc}({\bf k'})P_{ld}({\bf k'})-\frac{1}{2} P_{tl}({\bf k'})
P_{cd}({\bf k'}))\langle\tau^*_{ab}({\bf k})\tau_{cd}({\bf k'})\rangle \,,
\ea
where $P_{ij}=\delta_{ij}-\hat k_i\hat k_j$. 
Such convolutions can be written in terms of spectra as follows:
\ba
\langle\Pi^{*(S)}({\bf k})\Pi^{(S)}({\bf k'})\rangle &=& 
|\Pi^{(S)}(k)|^2\delta({\bf k}-{\bf k'})\nonumber\\
\langle\Pi_i^{*(V)}({\bf k})\Pi_j^{(V)}({\bf k'})\rangle &=& 
\frac{1}{2} |\Pi^{(V)}(k)|^2 P_{ij}({\bf k})\delta({\bf k}-{\bf k'})\nonumber\\
\langle\Pi_{ij}^{*(T)}({\bf k})\Pi_{tl}^{(T)}({\bf k'})\rangle &=& 
\frac{1}{4}|\Pi^{(T)}(k)|^2 \mathcal{M}_{ijtl} ({\bf k}) 
\delta({\bf k}-{\bf k'}) \nonumber 
\ea
where $\mathcal{M}_{ijtl}=P_{it}P_{jl}+P_{il}P_{jt}-P_{ij}P_{tl}$.
With this choice the spectra take the form:
\ba
|\rho_B(k)|^2&=&\frac{1}{1024\pi^5}\int d{\bf p} P_B(p) \, P_B(|{\bf k}-{\bf p}|)(1+\mu^2) 
\label{density}
\\
|\Pi^{(V)}(k)|^2&=&\frac{1}{512\pi^5}\int d{\bf p} P_B(p) 
\, P_B(|{\bf k}-{\bf p}|) \, \times \nonumber\\
&& [(1+\beta^2)(1-\gamma^2) + \gamma\beta(\mu-\gamma\beta)] 
\label{vector}
\\
|\Pi^{(T)}(k)|^2&=&\frac{1}{512\pi^5} \int d{\bf p} P_B(p) \, 
P_B(|{\bf k}-{\bf p}|) \, \times \nonumber\\
&& (1+2\gamma^2+\gamma^2\beta^2) \,, 
\label{tensor}
\ea
where $\mu = \hat {\bf p} \cdot ({\bf k} -{\bf p})/|{\bf k} -{\bf p}|$, 
$\gamma= \hat {\bf k} \cdot \hat {\bf p}$,
$\beta= \hat {\bf k} \cdot ({\bf k} -{\bf p})/|{\bf k} -{\bf p}|$. 
These equations agree, within our Fourier 
convention, with previous results by \cite{MKK}, \cite{DFK}.
One of the main results of our work is the calculation of the correct, 
i.e. without any approximation, expressions for these convolutions, 
given a power spectrum  as in Eq.(2.5) with a sharp cut-off at $k_D$. 

\begin{figure*}
%\centering
\begin{tabular}{cc}
\includegraphics[width=8cm]{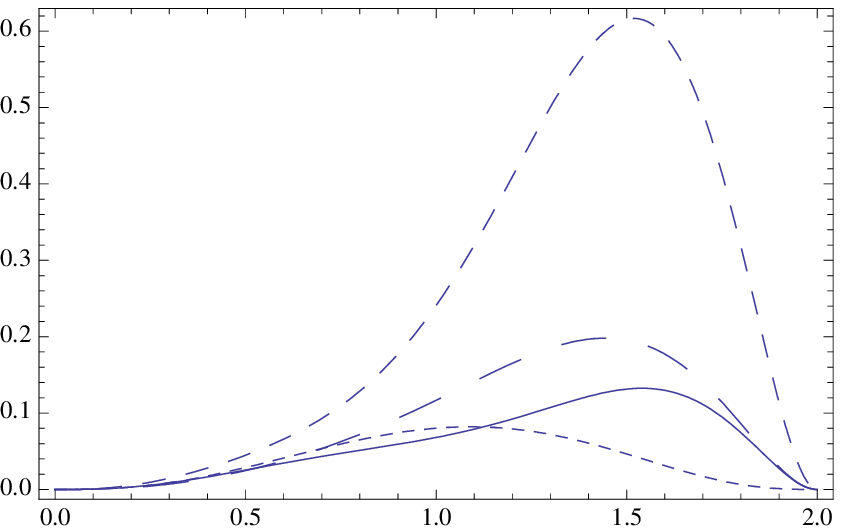}
\includegraphics[width=8cm]{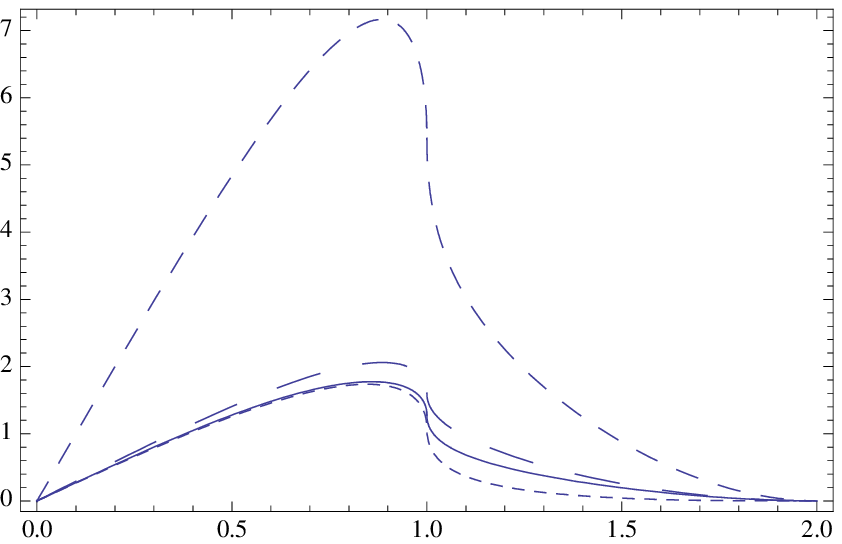}
\end{tabular}
\caption{Comparison of $k^3 |\rho_B(k)|^2$ (solid line), $k^3 |L (k)|^2$ (large dashed line)
$k^3 |\Pi_{i}^{(V)} (k)|^2$ (small dashed line), $k^3 |\Pi_{ij}^{(T)} (k)|^2$ (medium dashed line)
in units of $\langle B^2 \rangle^2/(1024 \pi^3)$
versus $k/k_D$. The left and right panel are for $n_B = 2$ and $n_B=-5/2$, respectively.}
\label{fig1}
\end{figure*}

In the appendices we explain the integration technique and 
show the results for various spectral indexes $n_B=3 \,, 2 \,, 1 \,,0 
\,, -1 \,, -3/2 \,, -5/2$: in this paper we add $n_B = -5/2$ to the 
previously studied $n_B \ge -3/2$ values for scalar quantities studied in FPP.
As for the scalar energy density and Lorentz 
force discussed in FPP, also the vector and tensor anisotropic stresses 
have support for Fourier modes with modulus smaller than $2 k_D$.
We show in Figure~\ref{fig1} the 
behaviour of scalar, vector and tensor quantities for $n_B=2, -5/2$, 
respectively. For $n_B = -5/2$ the spectra for $k \ll k_D$ is:
\be
|\rho (k)|^2_{n_B=-5/2} \simeq \frac{17 A^2 k_*^5}{800 \pi^4 k^2}
\ee 
whose slope, but not the amplitude, agrees with \cite{KR}.
%not white noise at $k \ll k_D$ and diverge as $(k/k_D)^{2 n_B+3}$; 
For $n_B = -5/2$, 
$| L(k) |^2 \simeq (55/51) |\rho (k)|^2$ for $k\ll k_D$ 
(for $n_B \ge -3/2$ we obtained $| L(k) |^2 \simeq (11/15) |\rho (k)|^2$ 
for $k\ll k_D$ FPP).  
Note also how the tensor contribution dominates over other ones
in amplitude of the Fourier spectra, in agreement with previous numerical 
findings \citep{browncrittenden}. 

\begin{figure*}
%\centering
\begin{tabular}{cc}
\includegraphics[width=8cm]{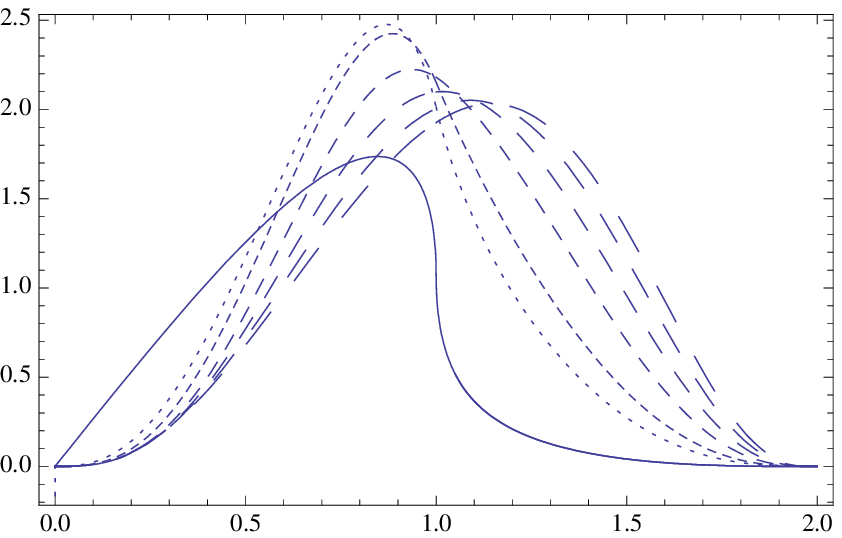}
\includegraphics[width=8cm]{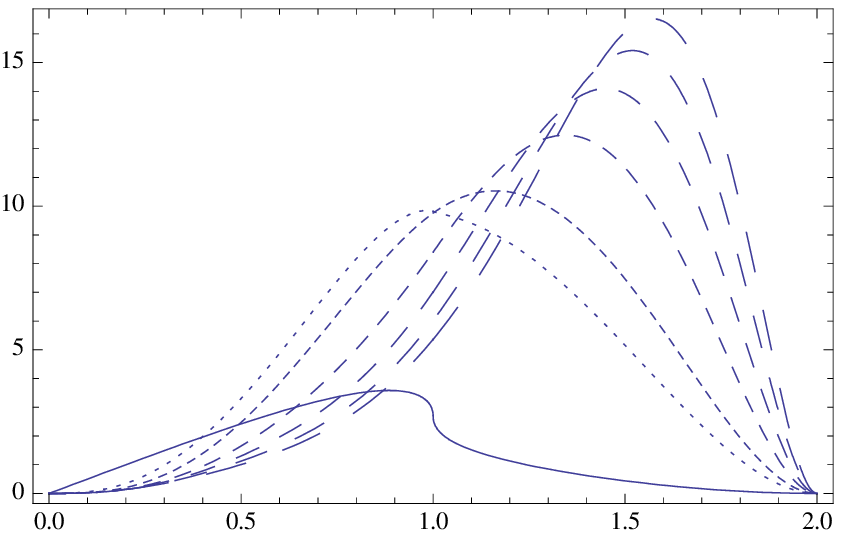}
\end{tabular}
\caption{Plot of $k^3 |\Pi^{(V)} (k)|^2$ (left panel) and 
$k^3 |\Pi^{(T)} (k)|^2$ 
(right panel) in units of $\langle B^2 \rangle^2/(1024 \pi^3)$ versus 
$k/k_D$ for different
$n_B$ for fixed $\langle B^2 \rangle$. The different lines are for
$n_B = -5/2, -3/2, -1, 0, 1, 2, 3$ ranging from the solid to the longest 
dashed.}
\label{fig2}
\end{figure*}

The vector and tensor anisotropic stresses are shown with 
varying $n_B$ in Figure~\ref{fig2}. The vector and tensor contributions have 
a $k$-dependence very similar to the energy-density: 
for $k \ll k_D$ and $n_B > -3/2$ ($n_B =-3/2$) 
$|\Pi^{(V)}(k)|^2 \,, |\Pi^{(T)}(k)|^2$ have a white noise (logarithmic 
divergent) spectrum with $|\Pi^{(T)}(k)|^2 \simeq 2 |\Pi^{(V)}(k)|^2$, 
whereas for $n_B = -5/2$ both become infrared dominated as 
$|\Pi^{(T)}(k)|^2 \simeq (94/25) |\Pi^{(V)}(k)|^2$ holds. 
The generic behaviour of $|\Pi^{(V)}(k)|^2$ for $k \ll k_D$ and $n_B > -3/2$ is:
\be
|\Pi^{(V)}(k)|^2 \simeq \frac{A^2k_D^{2n+3}}{256 \pi^4 k_*^{2n}
(3+2n_B)}\frac{28}{15} \,.
\label{infraredvector}
\ee
The pole for $n_B = -3/2$
in Eq. (\ref{infraredvector}) is replaced by a logarithmic diveregence in $k$
in the exact result reported in the Appendix A; the result reported by \cite{MKK} has a factor 4 instead of the factor $56/15$ reported in 
Eq. (\ref{infraredvector}).
Note also that the relation between the tensor and vector anisotropic stresses is different from
the one reported in \cite{MKK}, who predict 
(in our conventions): $|\Pi^{(T)}(k)|^2 = |\Pi^{(V)}(k)|^2$, this 
relation is obtained 
neglecting the angular part in 
Eqs.(\ref{vector},\ref{tensor}) and is incorrect.

%---------------------------------------------------------------------------- 

%---

%------------------------------------------------------------------------------------------------------------------

\section{Pertubations Evolution with PMF}

The presence of PMF influences the cosmological perturbations evolution mainly in three ways.
PMFs carry energy and momentum at the perturbation level and therefore gravitate, influencing the metric perturbations.
As a second point, they carry anisotropic stress, which adds to the ones already present in the plasma given by neutrinos and photons, with 
the caveat that the photon
anisotropic stress is negligible before decoupling epoch.
Third, the presence of PMF induces a Lorentz force on baryons, which modifies their velocity.
Due to the tight coupling between photons and baryons prior to the decoupling epoch the Lorentz force has an indirect 
effect also on photons during this regime.

The evolution of metric perturbations is described by the Einstein equations. These are modified by the presence of PMFs that 
represent a source term as follows:
\be
G_{\mu\nu}=8\pi (T_{\mu\nu}+\tau_{\mu\nu})\,,
\ee
where as usual $\tau_{\mu\nu}$ represents the PMF EMT. 
The metric chosen in this work is:
\be
ds^2 = a^2(\tau) \left[ - d\tau^2 + \left( \delta_{i j} + h_{i j} \right) 
d x^i d x^j \right]
\ee
%where the metric perturbations are splitted in scalar, vector and tensor parts:
where $h_{i j}$ can be decomposed into a trace part $h$ and a traceless part 
consisting of its scalar, vector and tensor part (Ma and Bertschinger 1995): 
\be
h_{i j} = \frac{h}{3} \delta_{i j} + \left( \partial_i \partial_j - 
\frac{\delta_{i j}}{3} \nabla^2 \right) \mu + h^{V}_{i j} + h^{T}_{i j} \,. 
\ee
The vector part being constructed in terms of a divergenceless vector 
$h_i^V$
\be 
h^{V}_{i j} = \partial_i h_j^V + \partial_j h_i^V \,. 
\ee
The tensor part is traceless and transverse ($\partial_i h^{T \, i}_{j}=0$).

\section{The Scalar Contribution}

We shall focus now on the magnetic scalar contribution to CMB anisotropies.
The effect on metric perturbations is described by the Einstein equations 
with a source term given by the PMF EMT.
We choose to work in the synchronous gauge where the 
scalar metric perturbation in the Fourier space is described by two scalar 
potentials, namely $h(k,\tau)$ and $\eta (k,\tau)$.
The Einstein equations with the contribution of PMF in the synchronous 
gauge are:
\begin{eqnarray}
&&k^2 \eta-\frac{1}{2} {\mathcal{H}} \dot h = - 4\pi
G a^2 (\Sigma_n\,\rho_n\delta_n+\rho_B)  \,,\nonumber \\
&&k^2 \dot\eta = 4\pi G a^2 \Sigma_n(\rho_n+P_n)\theta_n \,, \nonumber\\
&&\ddot h +2 {\mathcal{H}} \dot h -2 k^2 \eta = - 8\pi G a^2
(\Sigma_n \,c_{s \, n}^2\rho_n\delta_n  \nonumber \\
& &
+\frac{\delta\rho_B}{3}) \nonumber \,, \\
&&\ddot h+6\ddot \eta +2 {\mathcal{H}} (\dot h+6 \dot\eta)-2 k^2\eta
= -24 \pi G a^2 \times \nonumber \\
& & [\Sigma_n(\rho_n+P_n) \sigma_n+\sigma_B] ,
\label{Einsteineqs}
\end{eqnarray}
where $n$ represents the various species of the plasma, i.e. baryons, cold dark 
matter (CDM), photons and massless neutrinos.
The conservation of the PMF EMT  - $\nabla_\mu \tau^{\mu \, {\rm PMF}}_\nu=0$ - 
implies that $\rho_B({\bf x},\tau)=\rho_B({\bf x},\tau_0)/a(\tau)^4$ and 
the following relation between the magnetic anisotropic stress $\sigma_B$, 
the magnetic energy density $\rho_B$ and the Lorentz force $L$ holds:
\be
\sigma_B=\frac{\rho_B}{3}+L \,,
\label{stress_B}
\ee
Such Lorentz force modifies the Euler equation for baryons velocity, 
leading to observational 
signatures (see \cite{FPP_PRD} for the most recent discussion on this effect). 

\subsection{Initial Conditions}

The magnetized adiabatic mode initial conditions in the synchronous 
gauge deep in the radiation era are:
{\setlength\arraycolsep{2pt}
\begin{eqnarray}
h&=&C_1 k^2\tau^2 -\frac{C_1(5 + 4 R_\nu)}{36(15 + 4 R_\nu)} k^4 \tau^4 \nonumber \\
& & + \left[ - \frac{55 L_B}{336(15 + 4 R_\nu)}+ 
\frac{(-55 + 28 R_\nu)\Omega_B}{1008(15+4 R_\nu)} \right] k^4 \tau^4
\nonumber \\
\eta&=&2C_1-\frac{5+4R_\nu}{6(15+4R_\nu)}C_1 k^2\tau^2 + \nonumber \\
& & \left[ \frac{\Omega_B(-55+28 R_\nu)}{168(15+4R_\nu)}
-\frac{55 L_B}{56(15+4R_\nu)}\right]
k^2 \tau^2 \nonumber\\
\delta_\gamma&=&-\Omega_B-\frac{2}{3}C_1 k^2 \tau^2+\left[ 
\frac{\Omega_B}{6}+\frac{L_B}{2(1-R_\nu)}\right] k^2 \tau^2 \nonumber\\
\delta_\nu&=&-\Omega_B-\frac{2}{3} C_1 k^2 \tau^2
-\left[ \frac{\Omega_B(1-R_\nu)}{6R_\nu}+\frac{L_B}{2R_\nu}\right] k^2 \tau^2 
\nonumber\\
\delta_b&=&-\frac{3}{4}\Omega_B-\frac{C_1}{2} k^2 \tau^2+\left[ 
\frac{\Omega_B}{8}+\frac{3L_B}{8(1-R_\nu)}\right] k^2 \tau^2 \nonumber\\
\delta_c&=&-\frac{3}{4}\Omega_B-\frac{C_1}{2} k^2\tau^2\nonumber\\
\theta_\gamma&=&-\frac{C_1}{18}k^4\tau^3-\left[\frac{\Omega_B}{4}
+\frac{3}{4}\frac{L_B}{(1-R_\nu)} \right] k^2 \tau
\nonumber \\ 
& & + \left[ \frac{\Omega_B}{72}+
\frac{L_B}{24(1-R_\nu)}\right] k^4 \tau^3\nonumber\\
\theta_b&=&\theta_\gamma\nonumber\\
\theta_c&=&0\nonumber\\
\theta_\nu&=&-\frac{(23+4 R_\nu)}{18(15+4R_\nu)}C_1 k^4\tau^3 \nonumber \\
& & +\left[\frac{\Omega_B(1-R_\nu)}{4 R_\nu}+\frac{3}{4}\frac{L_B}{R_\nu}\right] 
k^2\tau \nonumber\\
& & -\left[\frac{(135+14 R_\nu)L_B}{84 R_\nu(15+4R_\nu)} \right. \nonumber \\
& & \left. -\frac{(-270+161 R_\nu+28 R_\nu^2)\Omega_B}{504 R_\nu (15+4R_\nu)}\right] k^4\tau^3 \nonumber\\
\sigma_\nu&=&\frac{4 C_1}{3(15+4R_\nu)} k^2\tau^2-\frac{\Omega_B}{4R_\nu}
-\frac{3}{4}\frac{L_B}{R_\nu} \nonumber \\
& & + \left[\frac{ -\Omega_B (-55+28 R_\nu)}{56 R_\nu(15+4R_\nu)}
+\frac{165 L_B}{56 R_\nu(15+4R_\nu)}\right] 
k^2\tau^2 \nonumber\\
F_{3\nu}&=&-\frac{6}{7} \left[\frac{\Omega_B}{4R_\nu}
+\frac{3}{4}\frac{L_B}{R_\nu} \right] k\tau\,,
\label{initialconds}
\end{eqnarray}}
where $\Omega_B = \rho_B/(\rho_\nu+\rho_\gamma)$, 
$L_B = L/(\rho_\nu+\rho_\gamma)$, 
$R_\nu=\rho_\nu/(\rho_\nu+\rho_\gamma)$, $C_1$ 
is the constant which characterizes 
the regular growing adiabatic mode as given in \cite{MB}.
The result reported is different from the one reported in
\cite{FPP_PRD} because we are adding the first
non-trivial terms in ${\cal O} (\Omega_B \,, L_B)$ for $h \,, \delta_c$ and
because of a different truncation in the neutrino hierarchy.
The term ${\cal O} (k^4 \tau^4)$ in $h$ (which also contains
the next-to-leading term of the adiabatic mode) has been obtained by
taking self-consistently the required order in all the variables in
Eqs. (\ref{initialconds}). For simplicity we have written
the ${\cal O} (k^4 \tau^4)$ term only in $h$ (since it is the leading term
for the magnetic solution) and omitted these higher order
terms in all the other variables.
Here we also choose to truncate the hierarchy at $F_{4\nu}=0$ instead 
of $F_{3\nu}=0$ as in \cite{FPP_PRD}.
This change affects the magnetic next to leading order terms in the 
velocity and anisotropic stress of neutrinos and in the metric 
perturbation $\eta$.
The equation for the evolution of the $F_{3\nu}$ is:
\be
\dot F_{3\nu}=\frac{6}{7}k \sigma_\nu\,,
\ee
while the neutrino anisotropic stress equation becomes:
\be
2\dot\sigma_\nu=\frac{8}{15}\theta_\nu-\frac{3}{5}k F_{3\nu}+\frac{4}{15}\dot h+\frac{8}{5}\dot\eta\,.
\ee

Note how the presence of a SB of PMFs induces a new independent mode
in matter and metric perturbations, i.e. the fully magnetic mode obtained
by setting $C_1=0$ in Eq. (\ref{initialconds}).
This new independent mode is the particular solution of the 
inhomogeneous system of the Einstein-Botzmann 
differential equations: the SB of PMF treated as a stiff source acts indeed 
as a force term in the system of linear differential equations. 
Whereas the sum of the fully magnetic mode with the curvature one can be with 
any correlation as for an isocurvature mode, the nature of the fully 
magnetic mode - and therefore its effect - is different: the isocurvature 
modes are solutions of the homogeneous system (in which all the 
species have both background and perturbations), whereas the 
fully magnetic one is the solution of the inhomogeneous system sourced 
by a fully inhomogeneous component.

It is interesting to note how the magnetic contribution drops from the 
metric perturbation at leading order, although is larger than the 
adiabatic solution for photons, neutrinos and baryons. 
This is due to a 
compensation which nullifies the sum of the leading contributions (in 
the long-wavelength expansion) in the single species energy densities 
and therefore in the metric perturbations. 
A similar compensation exists for a network of topological defects, 
which does not carry a background energy-momentum tensor as the PMF SB 
studied here\footnote{Note however that a network of topological defects 
does not scale with radiation and 
interacts only gravitationally with the rest of matter, i.e. a 
Lorentz term is absent.}. 

\subsection{Analytic Description of the Scalar Magnetic Contribution 
on Large Angular Scales}

In this subsection we give an analytic description 
for the scalar magnetic contribution to CMB anisotropies on large angular 
scales given by the initial conditions in 
Eq. (\ref{initialconds}) and computed by our modified version of CAMB. 
The Sachs-Wolfe term is 
$\delta T/T_{| {\rm SW}} \sim \delta_r/4 + \psi$, where $\delta_r$ is the
radiation density contrast and $\psi$ is one of the metric
potentials in the longitudinal gauge:
\begin{equation}
ds^2 = a^2 (\eta) \left[ - (1 + 2\psi) d\eta^2 + (1-2\phi) \gamma_{ij}
d x^i d x^j \right]
\end{equation}
For adiabatic initial conditions is simply 
$\delta T/T_{| {\rm SW}} \sim \psi/3$ since
$\delta_r \simeq - 8 \psi/3$ in the matter dominated era on large scales.

The Sachs-Wolfe term for the scalar magnetic mode generated by a
SB of PMF can be obtained by using the initial conditions given in the
synchronous gauge in Eq. (\ref{initialconds}). 
By making a gauge transformation, we obtain at leading order in the 
radiation era:
\begin{eqnarray}
\frac{\delta_r}{4} &\simeq& - \frac{\Omega_B}{4}
+ \frac{\Omega_B(1-R_\nu)+3L_B}{15+4 R_\nu} \nonumber \\
\psi &\simeq& - \frac{\Omega_B (55-28 R_\nu)+165 L_B}{14(15+4 R_\nu)} \,.
\end{eqnarray}
%Although the above initial conditions are in the radiation
%era and not in the matter era, these are very useful to show that,
In the radiation era, 
because of compensation, the metric potential are just proportional to
$\Omega_B$ and not to $\Omega_B/k^2$ as obtained by \cite{KR}. As we 
will show in the following, the same holds in the matter era.

Now let us assume that in the matter era:
\begin{equation}
\frac{\delta T}{T}_{| {\rm SW}} = \alpha \frac{\Omega_B}{4}
\end{equation}
and we compute the scalar contribution to CMB anisotropies for $n > -3/2$ 
by solving the integral:
\begin{equation}
C_\ell^{S \,, magnetic} \simeq \frac{\alpha^2}{8 \pi} \int_0^{k_D} dk k^2 |\Omega_B|^2
j^2_\ell (k \eta_0) \,,
\end{equation}
where we have insterted an upper cut-off $k_D$ in order to use the 
infrared expansion of $|\rho_B(k)|$, obtained in FPP:
\be
|\rho_B(k)|^2 \simeq \frac{A^2k_D^{2n+3}}{128 \pi^4 k_*^{2n} (3+2n_B)} \,.
\label{infrareden}
\ee

%With the notation of arXiv:0803.1246 we obtain 
By using the result \citep{AS}
\be
\int_0^y dx x J^2_{\ell + 1/2} (x) = \frac{y^2}{2} \left[
J^2_{\ell + 1/2} (y) - J_{\ell - 1/2} (y) J_{\ell + 3/2} (y) \right] 
\nonumber \,,
%\simeq \frac{y}{\pi} \,.
\ee
we obtain for $n > -3/2$ and $y >> 1$:
\begin{equation}
C_\ell^{S \,, magnetic} \simeq \frac{\alpha^2 (n+3)^2 \langle B^2 \rangle^2}
{512 \pi (2n+3) \rho_{REL \,, 0}^2 k_D^2 \eta_0^2} \,.
\label{sw_analytic}
\end{equation}
The scalar magnetic contribution to CMB anisotropies on large angular scales 
is therefore white noise 
($C_\ell^{S \,, magnetic} \propto \ell^2$) for $\ell < 400$ and $n_B > -3/2$; 
the slope in $\ell$ of Eq. (\ref{sw_analytic}) and of the numerical results 
obtained with our modified version of CAMB agree very well. 
By using an analogous procedure for $n_B = -5/2$, we obtain analitically 
$C_\ell^{S \,, magnetic} \sim 1/\ell$ to be compared with the numerical 
results -$\sim 1/\ell^{0.4}$- obtained with our modified version of CAMB. 
The $\alpha$ parameter in Eq. (\ref{sw_analytic}) 
we can fit from our numerical results inherits 
a dependence on $n_B \,, k_D$ and is tipically $\sim {\cal O} (0.01 - 0.1)$.

\section{The Vector Contribution}

In this section we shall review the evolution of vector perturbations 
induced by a SB of PMF as treated in \cite{lewis}.
%Vector perturbations are created by vorticity, which decays rapidly in an isotropic and homogeneous  
%universe. 
%This leads to a suddenly disappeareance of an hypotetically inflationary vector mode.
%Therefore magnetically driven vector modes represent a unique signature of PMF on CMB and can affect CMB 
%anisotropies both in temperature and polarization.
The vector metric perturbation is described through a divergenceless vector:
\be
h^{V}_ {ij} = \partial_i h_j + \partial_j h_i
\ee
where we have
\be
\partial_i h_i=0
\ee
The divergenceless condition assures that vector mode does not support density perturbations.
The Einstein equations in the presence of PMF for the vector metric perturbations simply reduce to: 
\be
\dot{h}^{V}+2{\mathcal{H}}h^{V}=- 16 \pi G a^2 (\Pi^{(V)}_\nu+\Pi^{(V)}_\gamma+\Pi^{(V)}_B)/k
\ee
Conservation equations for PMF also in the vector case reduce to a relation between isotropic and anisotropic  pressure and the vector Lorentz force:  
\be
-\nabla_i p^B+ \nabla_j\Pi_{ij}^{(V)B}=L_i^B
\ee
The Lorentz force induced on baryons, in analogy with what 
we found for the scalar case, modifies the baryon vector velocity equation:
\be
\dot v_b+{\mathcal{H}}v_b=-\frac{\rho_\gamma}{\rho_b}
\Big(\frac{4}{3} n_e a \sigma_T(v_b-v_\gamma)
-\frac{L^V}{\rho_\gamma}\Big)
\ee 
where we have neglected the baryon homogeneous pressure ($p_b/\rho_b<<1$). 
In order to investigate the effect of magnetized vector perturbations it is necessary to calculate the Fourier spectra  for the
vector projection of the PMF EMT and the Lorentz force.
Since these two quantities are related by:
\be
L_i^{(V)}=k \Pi_i^{(V)}\,,
\ee 
we need only one spectrum to compute for the vector part, as for the tensor 
part described in the next section and differently from the scalar part.  
%and therefore contrary to what we had for the scalar case, 
%vector mode requires only one spectrum which is the vector 
%projection of the PMF EMT:
%{\setlength\arraycolsep{2pt}
%\ba
%<\Pi_i^{(V)}(k)\Pi_i^{(V)*}(k)>&=&\frac{1}{16\pi^2}\frac{1}{(2\pi)^3}\int d {\bf p} P(p)P(|{\bf k}-{\bf p}|)\times\nonumber\\
%&&[(1-\gamma^2)(1+\beta^2)+\gamma\beta(\mu-\gamma\beta)]\nonumber\\
%\ea
%}
%where we have used the same notation than in the scalar section.
%In the Appendix we present the results of this convolution for 
%several spectral indexes.

\section{The Tensor Contribution}
Inflationary tensor modes, namely primordial gravitational waves, are a 
key prediction of the standard inflationary model, and therefore their 
indirect observation through
CMB anisotropies is one of the crucial point of modern cosmology.
However PMF carrying anisotropic stress are themselves a source of tensor 
perturbations. Therefore
the presence of PMF affects inflationary tensor modes and creates a new independent fully magnetic tensor mode in analogy to what we found for scalar perturbations.
The evolution equation for the metric tensor perturbation $h_{ij}$ is:
\be
\ddot{h}_{ij}+2{\mathcal{H}}\dot{h}_{ij}+ k^2 h_{ij}=16\pi G a^2(\rho_\nu\pi^\nu_{ij}+\Pi^{(B,T)}_{ij})\,.
\ee
which, for each polarization state deep in radiation era reads
\be
\ddot{h}^T_k+\frac{2}{\tau}\dot{h}^T_k+ k^2 h^T_k=\frac{6}{\tau^2}[R_{\nu}\sigma^{(T)}_\nu+(1-R_{\nu})\tilde\Pi^{(T)}_B]\,
\label{tensor_evol}
\ee
where $\tilde\Pi^{(T)}_B$ represents the time independent variable 
$\Pi^{(T)}_B/\rho_\gamma$.
The large scales solution to this equation can be found expanding $h_k$ in powers of $(k\tau)$. In order to keep the leading and the next-to-leading terms we need to take into account the neutrino octopole $(J_3)$, truncating the propagation of anisotropic stress through higher moments by posing $J_4=0$. Hence the neutrino anisotropic stress evolves according to
\begin{eqnarray}
\dot{\sigma}^{(T)}_\nu &=& -\frac{4}{15}\dot{h}_k-\frac{k}{3}J_3 \nonumber\\
\dot{J}_3 &=& \frac{3}{7}k\sigma^{(T)}_\nu
\end{eqnarray}
and the solution is then:
\begin{eqnarray}
h_k &=& A\Big[1-\frac{5(k\tau)^2}{2(15+4R_\nu)}\Big]+\frac{15 (1-R_\nu)\tilde\Pi^{(T)}_B (k\tau)^2}{14(15+4R_\nu)} \,,\nonumber\\
\sigma^{(T)}_\nu &=&-\frac{(1-R_\nu)}{R_\nu}\tilde\Pi^{(T)}_B\Big[1-\frac{15(k\tau)^2}{14(15+4R_\nu)}\Big]\nonumber \\
& & +A\frac{2(k\tau)^2}{3(15+4R_\nu)}\,.
\end{eqnarray}

The presence of magnetic fields is responsible for the new leading 
term in $\sigma^{(T)}_\nu$ - otherwise absent. This is the so-called 
compensation between collisionless fluid and magnetic anisotropic stresses 
due to fact that magnetic fields gravitate only at perturbative level.

Note that the compensation between anisotropic stresses
turns on only after neutrino decoupling,
an epoch which is much earlier than the usual initial time at which
cosmological perturbations are evolved in an Einstein-Boltzmann code.
The detailed study of the evolution of gravitational waves during neutrino
decoupling is an interesting issue, but clearly
beyond the purpose of the present project.
  
%Two new next-to-leading terms appear as well in $\sigma^{(T)}_\nu$ and $h_k$ respectively.

%In order to study the magnetic tensor mode we need thus to calculate 
%the tensor part of EMT:
%\be
%|\Pi_B^{(T)}(k)|^2=\frac{1}{64\pi^2}\frac{1}{(2\pi)^3}\int d {\bf p} 
%P(p)P(|{\bf k}-{\bf p}|)[(1+2\gamma^2+\gamma^2\beta^2)]
%\ee
%where the notation is the same as in the scalar and vector section.
%Again in the Appendix we report the results of this calculation. 
%\label{tensor}

\section{Results for CMB anisotropies}

In this section we now present the temperature and polarization CMB
spectra including the full contribution of SB of PMF.
In addition to the $C_\ell$ obtained by the adiabatic mode in absence of
primordial magnetic fields, we add the
three contributions $C_\ell^{\rm S,V,T}$ described in Sects. 4,5,6, i.e.
scalar, vector, tensor, respectively, computed separately
by our modified version of CAMB. In Fig.3 and Fig.4 we show the results
for $n_B=2$ and $-5/2$, respectively. For the initial conditions of the scalar
magnetic mode, we use as initial conditions Eq. (4.22) with $C_1=0$; 
for those of the tensor mode the ones in Eq. (6.40) with $A=0$.
For the initial conditions for the vector mode we use the ones already
implemented in CAMB, described in \cite{lewis}.
All the formulae for magnetic spectra needed - $\Omega_B \,, L_B \,, \Pi_V,
\Pi_T$ - are given in our appendices; the signs of $L_B$ and
$\Omega_B$ are taken as opposite, as explained in \cite{FPP_PRD}.

\begin{figure*}
%\centering
\begin{tabular}{cc}
\includegraphics[width=8cm]{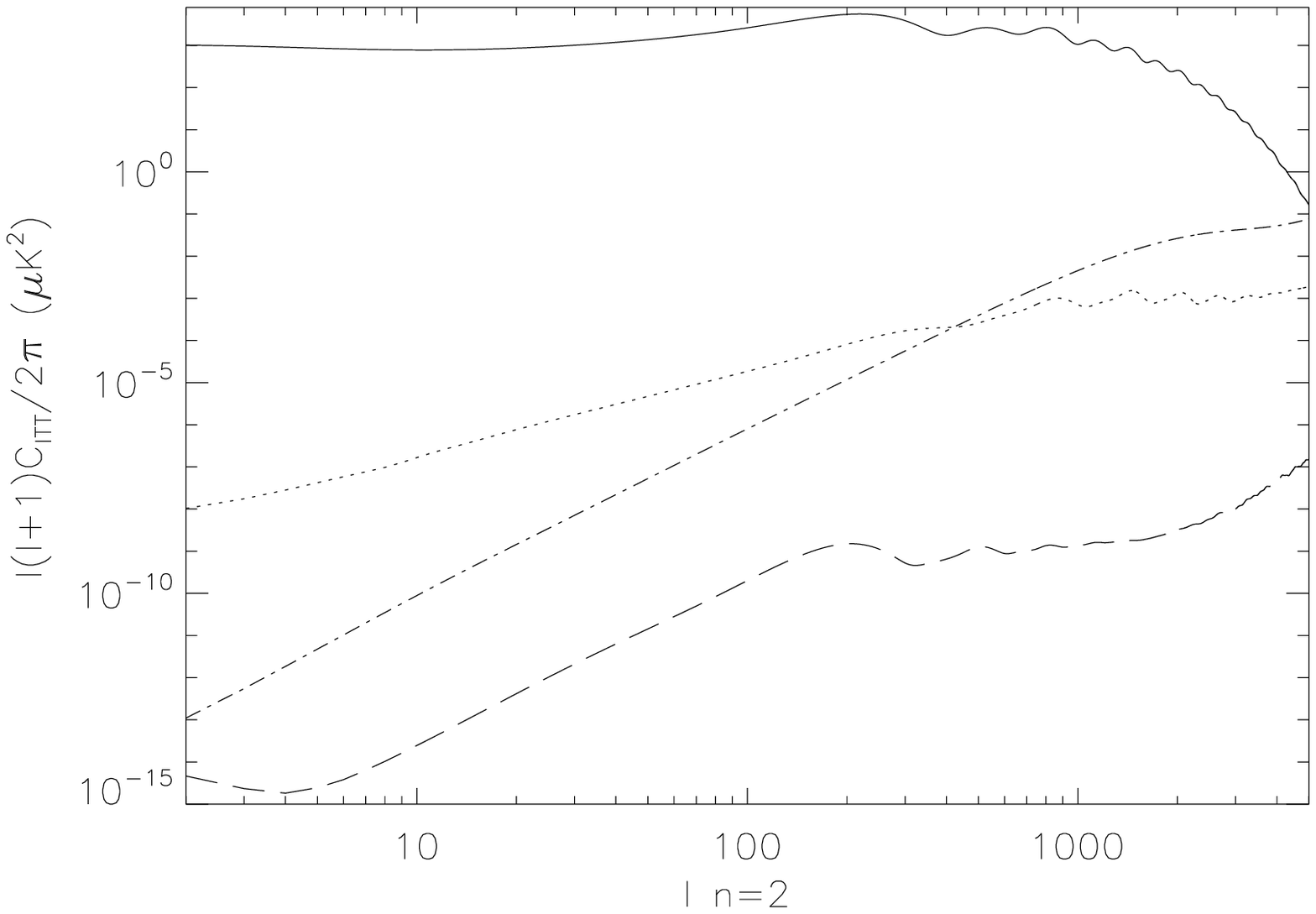}
\includegraphics[width=8cm]{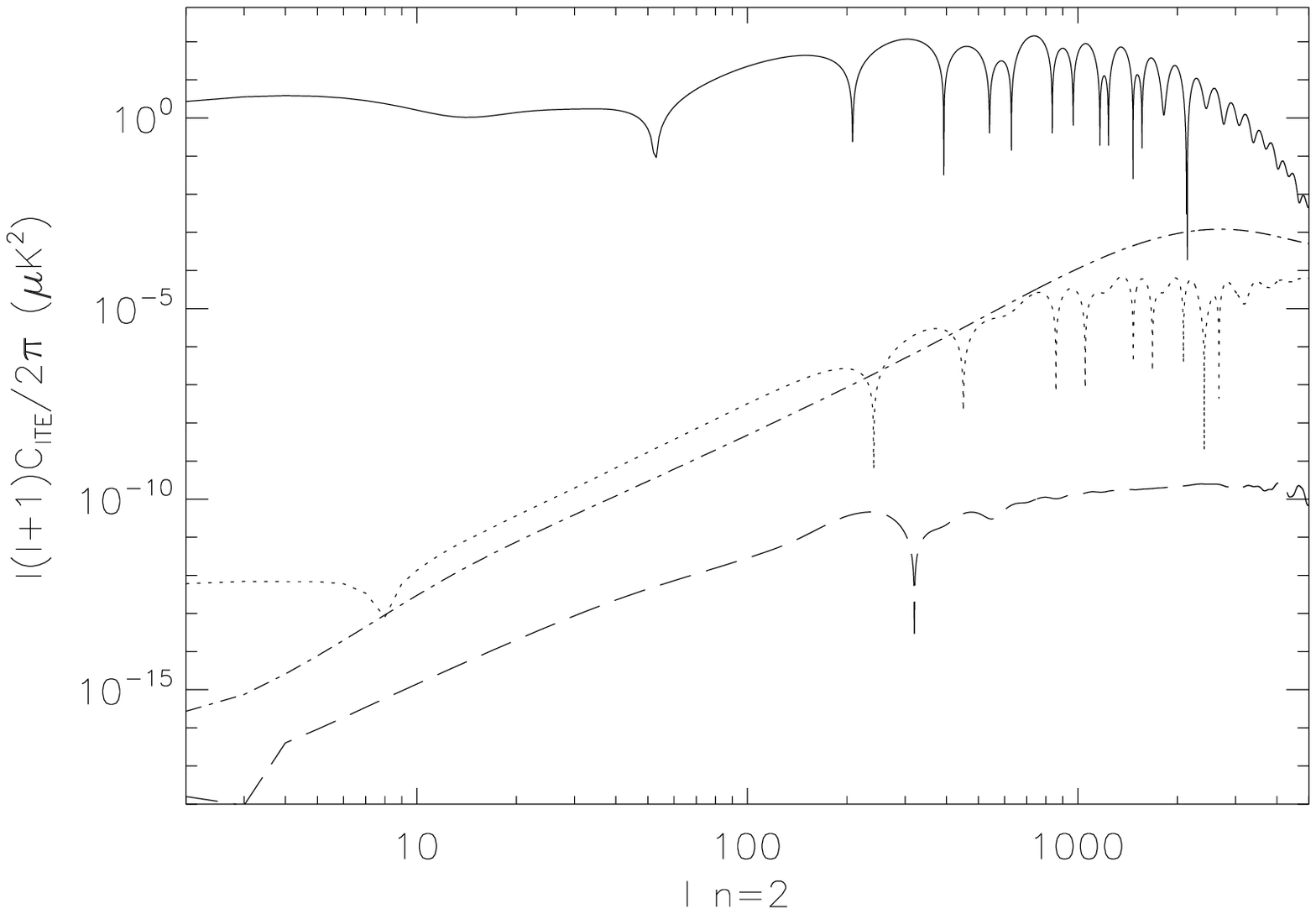} \\
\includegraphics[width=8cm]{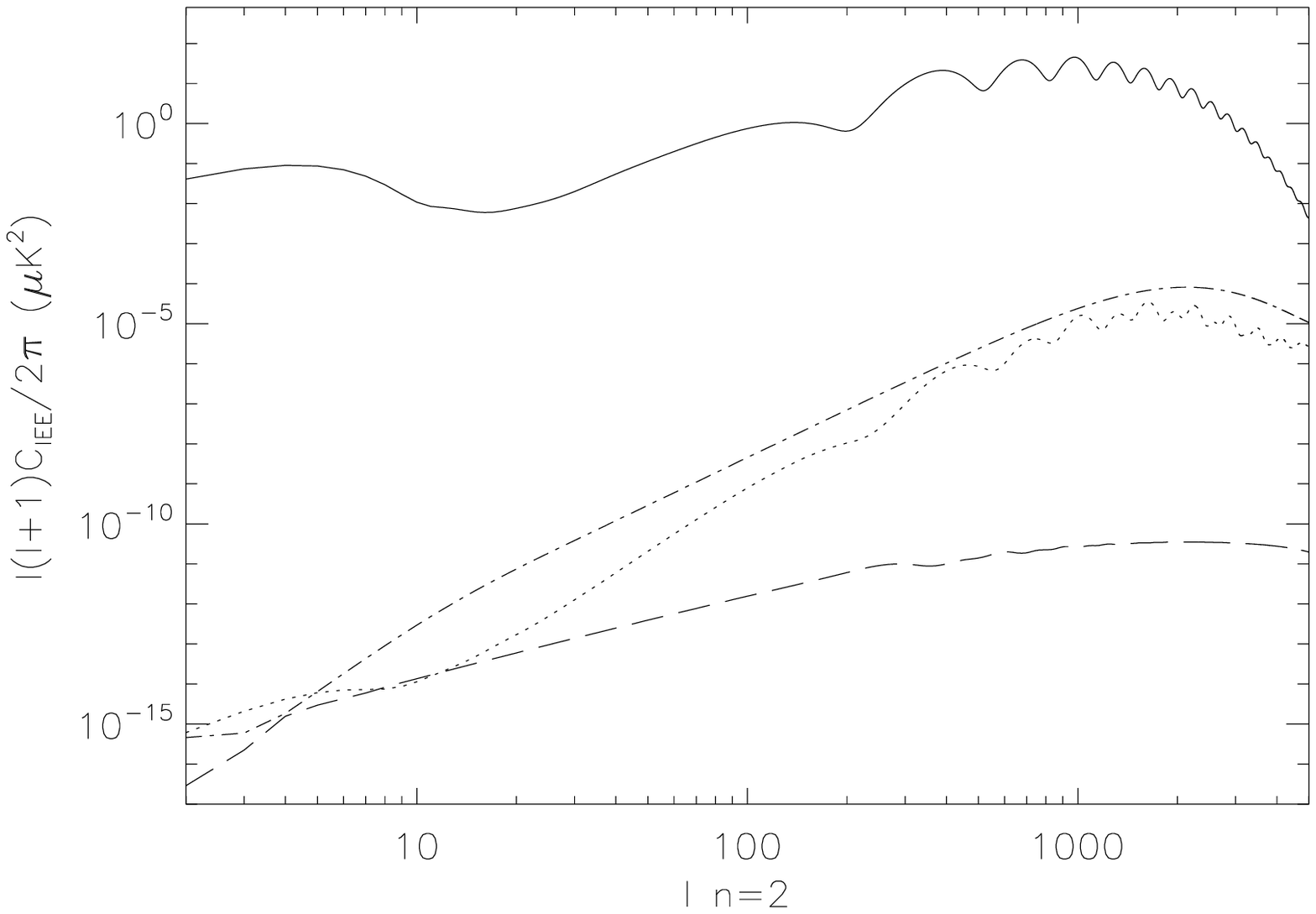}
\includegraphics[width=8cm]{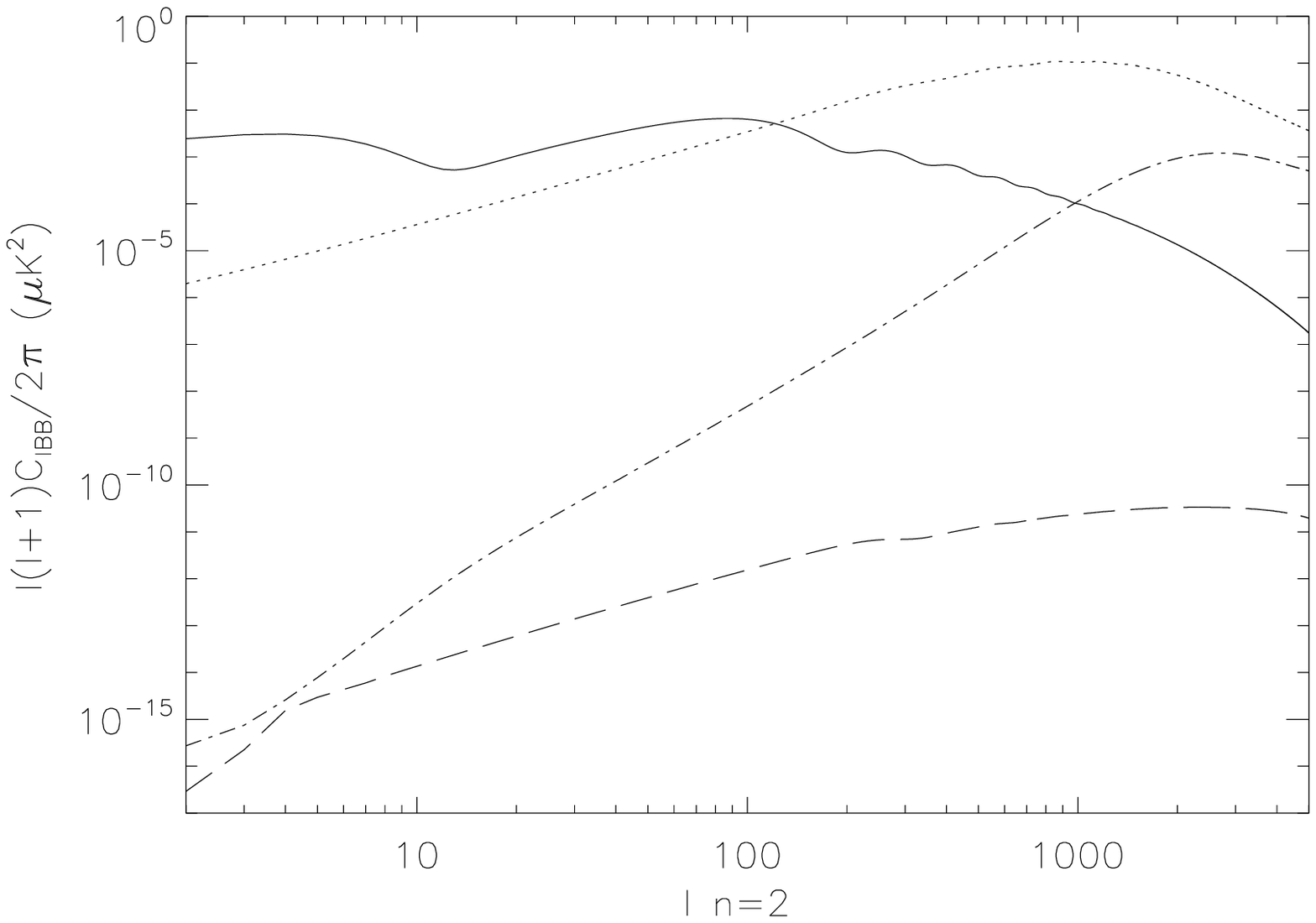}
\end{tabular}
\caption{
CMB anisotropies angular power spectrum for temperature (TT hereafter,
top-left panel),
temperature-E polarization cross correlation (TE hereafter, top-right
panel),
E polarization (EE hereafter, bottom-left panel), B polarization (BB
hereafter, bottom-right panel). 
The solid line is the adiabatic scalar contribution in TT, TE, EE panels, 
whereas it is the tensor homogeneous contribution in the BB panel 
(for a tensor-to-scalar ratio $r =0.1$);  
the dotted, dot-dashed, dashed are the scalar, vector and tensor contributions 
of a SB of PMF respectively for $\sqrt{\langle B^2 \rangle} = 7.5 \,$ nG, 
$n_B = 2$ and $k_D = 2 \pi \, {\rm Mpc}^{-1}$. The dotted line in 
the BB panel is the lensing contribution. 
%The damping scale $k_D$ has been fixed to $ $. 
The cosmological parameters of the flat $\Lambda CDM$ model are
$\Omega_b \, h^2 = 0.022$,
$\Omega_c \, h^2 = 0.123$, $z_{\rm re}= 12$, $n_s=1$,
$H_0 = 100 \, h \, {\rm km \, s}^{-1} \, {\rm Mpc}^{-1} 
= 72 \, {\rm km \, s}^{-1} \, {\rm Mpc}^{-1}$.}
\label{fig3}
\end{figure*}

\begin{figure*}
%\centering
\begin{tabular}{cc}
\includegraphics[width=8cm]{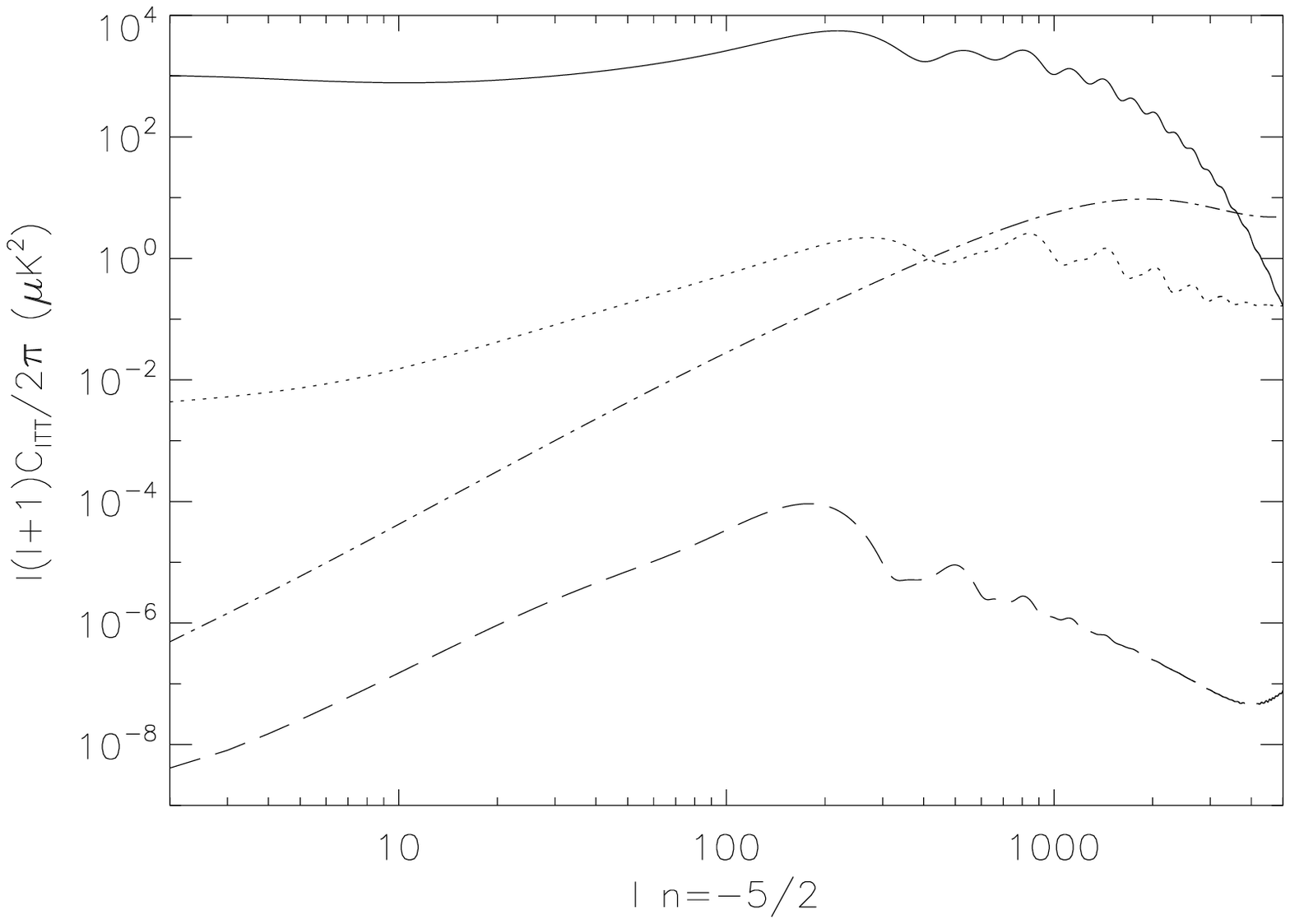}
\includegraphics[width=8cm]{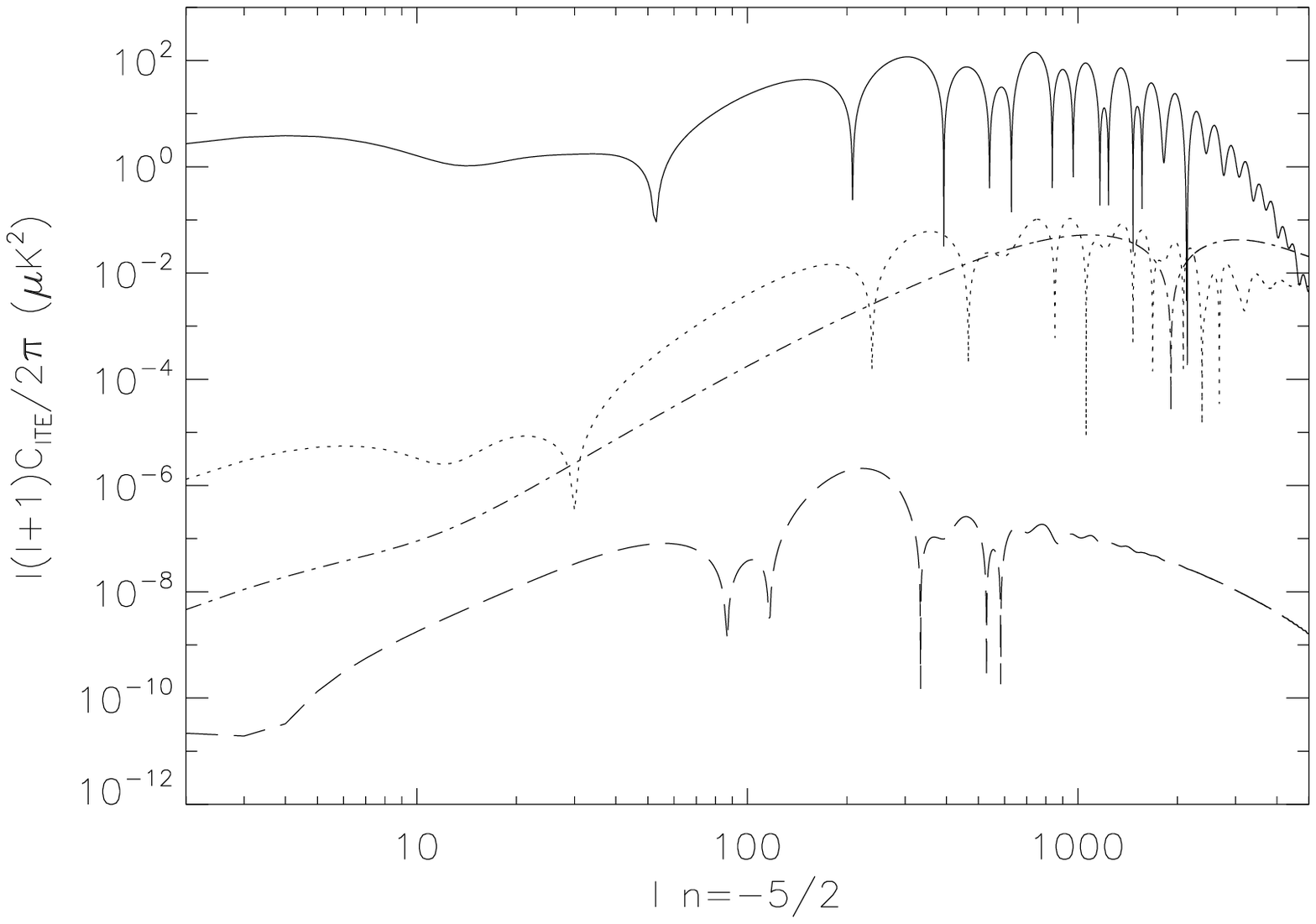} \\
\includegraphics[width=8cm]{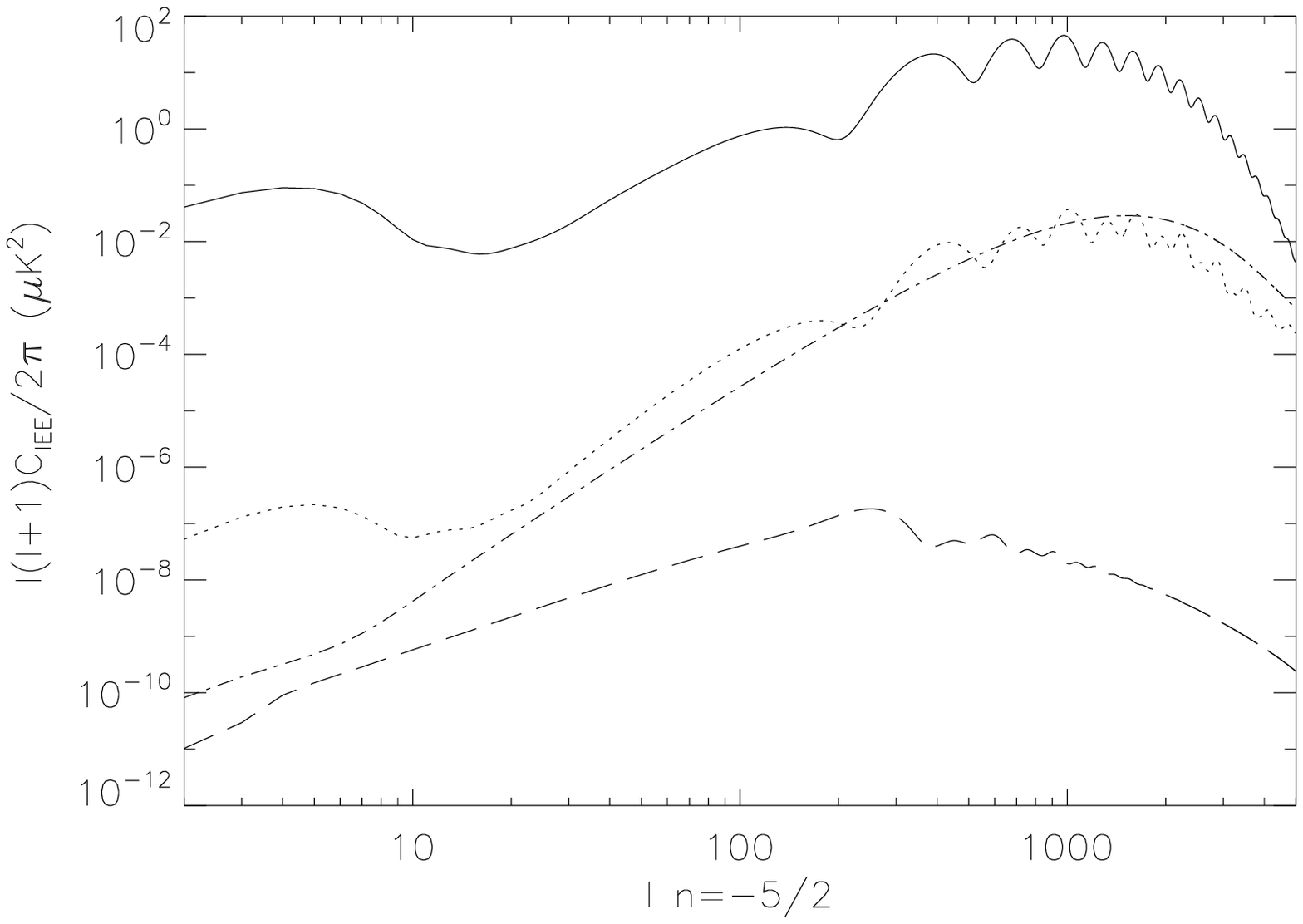}
\includegraphics[width=8cm]{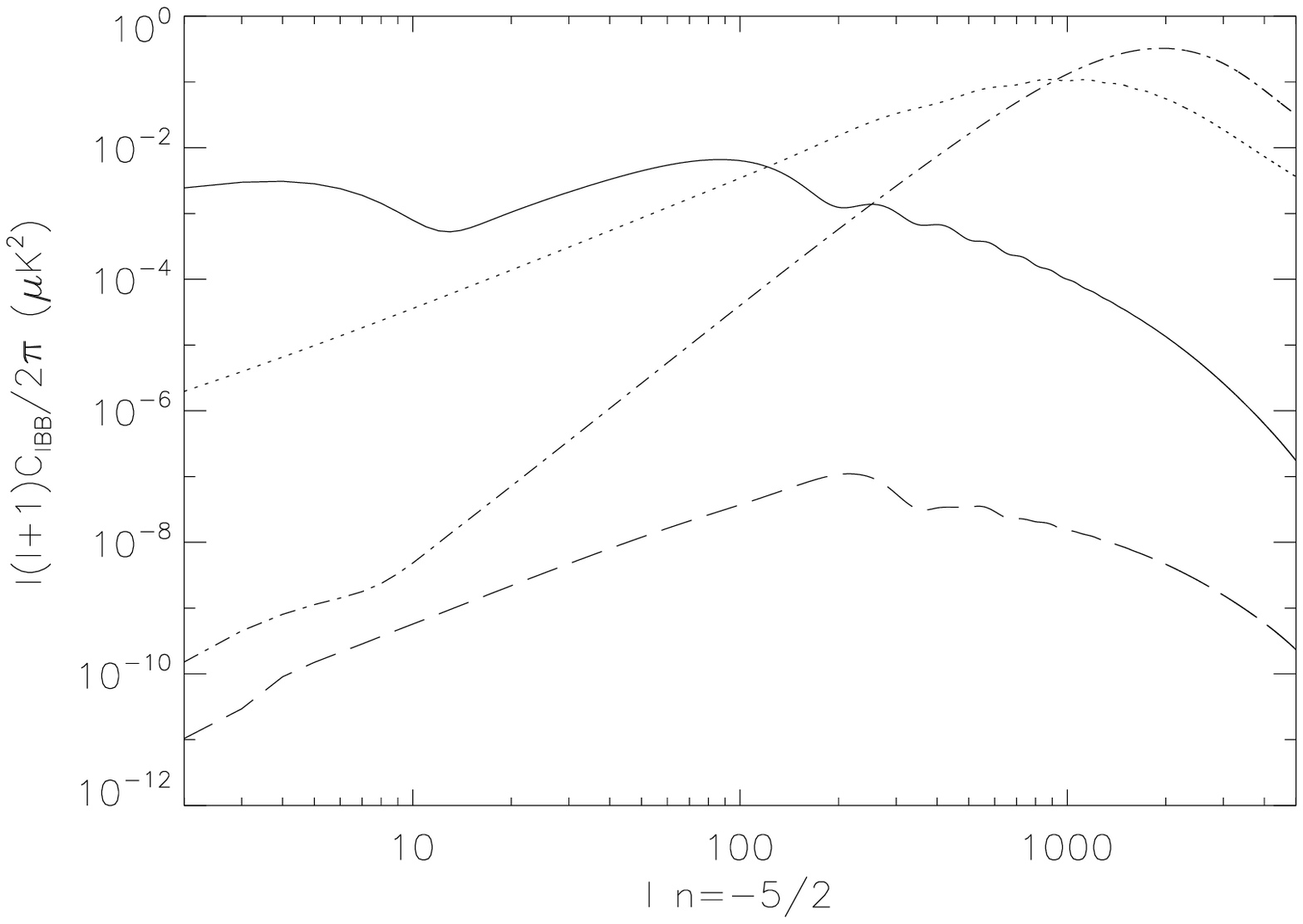}
\end{tabular}
\caption{
CMB angular power spectrum for TT (left top panel), TE (left top panel),
EE (bottom left), BB (bottom right).
The solid line is the adiabatic scalar contribution in TT, TE, EE panels,
whereas it is the tensor homogeneous contribution in the BB panel (for a tensor-to-scalar ratio $r =0.1$);
the dotted, dot-dashed, dashed are the scalar, vector and tensor contributions
of a SB of PMF respectively for $\sqrt{\langle B^2 \rangle} = 7.5 \,$ nG, 
$n_B = -5/2$ and $k_D = 2 \pi \, {\rm Mpc}^{-1}$. The dotted line in 
the BB panel is the lensing contribution.
%The damping scale $k_D$ has been fixed to $ $.
The cosmological parameters of the flat $\Lambda CDM$ model are the same as in Fig. 3.}
\label{fig4}
\end{figure*}

For all the values of $n_B$ considered here, the 
CMB temperature pattern generated by the
SB of PMF is dominated by the scalar contribution at low and 
intermediate multipoles; the vector contribution takes over the scalar one 
at high multipoles, whereas the tensor one is always subleading with 
respect to scalar and vector. 

It is interesting to note that the $B$ polarization signal due to the 
vector contribution is always 
larger than the tensor one. The $B$ mode produced by vector  perturbations
has a power spectrum which can be steeper than the one produced by lensing 
with a peak around $\ell \sim {\rm few} \times 10^3$; 
therefore, for suitable values of the magnetic field amplitude the 
$B$ mode produced by a SB of PMF can be larger than the lensing one 
for any $n_B$. Fig. 5 shows how the vector contribution to the 
$B$ spectrum depends on $n_B$. For $n_B > - 3/2$ the 
$B$ spectra from the vector contribution are almost 
indistinguishable for different $n_B$, because $\Pi^{(V)}_B$ is white noise 
for $k \ll k_D$; for $-3 < n_B \le -3/2$ we note a dependence of 
the $B$ spectrum on $n_B$. Analogous dependence on $n_B$ also holds for the vector contribution to $TT$.

\begin{figure*}
\includegraphics[width=8cm]{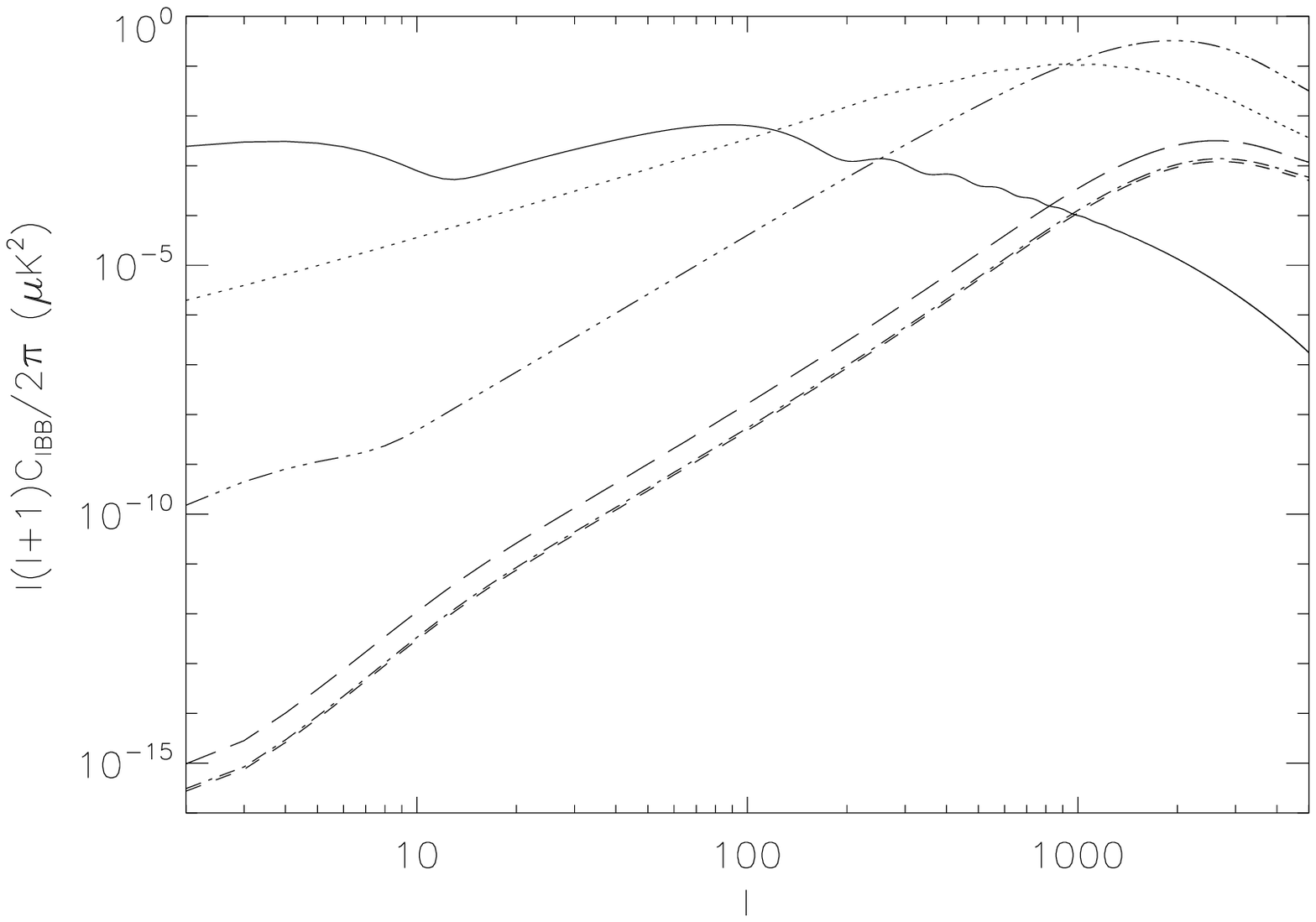}
\caption{Vector contributions to the CMB angular power spectrum for BB.
The solid line is the tensor homogeneous contribution for a tensor-to-scalar ratio $r =0.1$ and the 
dotted line is the lensing contribution with cosmological parameters as in the previous figures;
the triple-dotted, long dashed, dot-dashed and dashed are the vector spectra obtained with 
$\sqrt{\langle B^2 \rangle} = 7.5 \,$ nG, $k_D = 2 \pi {\rm Mpc}^{-1}$, 
for $n_B = -2.5, -1.5, -1, 2$, respectively. Note how the spectra for $n_B=-1$ and $n_B=2$ are 
super-imposed since the Fourier spectra of the vector part of the PMF EMT are both white noise 
for $k \ll k_D$ for these spectral indexes.}
\label{fig5}
\end{figure*}

\section{Conclusion}

We have obtained the Fourier spectra of the relevant scalar, 
vector and tensor energy-momentum components of the SB of PMF 
extending the method 
used in \cite{FPP_PRD} only for the scalar.
As already discussed for the scalar sector in \cite{FPP_PRD}, 
we have shown how the correct evaluation of 
the convolution integrals leads to differences in the vector 
and tensor parts of the PMF Fourier spectrum 
previously found in \cite{MKK}. 

We have then shown the comparison of the scalar, 
vector and tensor contributions to CMB anisotropies of the new 
inhomogeneous modes generated by the SB of PMF, by using the correct 
convolutions 
for the energy-momentum tensor of the PMF SB. 
We have shown that the dominant contributions are from scalar and vector 
perturbations, respectively for low and high $\ell$.
We have given an analytic description of the Sachs-Wolfe contribution of the  
scalar mode, which agrees very well with the numerical result of our 
modified version of CAMB \citep{CAMB} and takes into account the 
compensation effect on large scales which is generic when a fully 
inhomogeneous source is present.
The slope in $\ell$ of the vector power spectrum we obtain numerically 
agrees very well with previous analytic \citep{MKK} and 
numerical \citep{lewis} results. As already found in \citep{lewis}, the $B$ 
signal by vector perturbations, with a peak slightly dependent on $n_B$ around 
$\ell \sim 2000$, has a shape different either 
from the inflationary gravitational waves or the lensing signal. 
We have characterized its dependence on $n_B$ and shown how this is 
non trivial for $n_B \le -3/2$: such signal can be constrained by Planck 
\citep{bluebook} and future small scale CMB polarization experiments.

%\acknowledgements

\appendix
\begin{onecolumn}
\section{EMT fourier spectra}

We use the convolutions for the PMF EMT  spectra 
with the parametrization for the magnetic field power spectrum given in Eq. (\ref{PSpectrum}). 
Since $P_B (k) =0$ for $k > k_D$,
two conditions need to be taken into account:
\be
p<k_D\,, \qquad \qquad \qquad | {\bf k}-{\bf p} | <k_D \,.
\ee
The second condition introduces 
a $k$-dependence on the angular integration domain and the two 
allow the energy power spectrum to be non zero only for $0<k<2k_D$. 
Such conditions split the double integral (over $\gamma$ and over $p$) 
in three parts
depending on the $\gamma$ and $p$ lower and upper limit of integration. 
For simplicity we normalize
the Fourier wavenumber to $k_D$ and we show the integrals with this convention.
%\begin{widetext}
A sketch of the integration is thus the following:
\begin{eqnarray}
1)&&  0<k<1 \nonumber\\
&&\int_{0}^{1-k}dp 
\int_{-1}^{1}d\gamma\,\dots + \int_{1-k}^{1}dp 
\int_{\frac{k^2+p^2-1}{2kp}}^{1}d\gamma\,\dots \equiv 
\int_{0}^{1-k}dp I_a (p,k) + \int_{1-k}^{1}dp I_b (p,k) 
\nonumber\\
2) && 1<k<2 \nonumber\\
&& \int_{k-1}^{1}dp 
\int_{\frac{k^2+p^2-1}{2kp}}^{1} d\gamma\,\dots \equiv 
\int_{k-1}^{1}dp I_c (p,k)
\label{intscheme}
\end{eqnarray}
%\begin{figure}
%\includegraphics[scale=0.54]{kpsplitting.eps}
%\caption{Integration domains in $(k,p)$ plane}
%\label{kpsplitting}
%\end{figure}

Particular care must be used in the radial integrals. 
In particular, the presence of the term $|k-p|^{n+2}$ in both integrands, 
needs a further splitting of the integral domain for odd $n$:\\
\begin{displaymath}
\int_0^{(1-k)} dp\rightarrow
\left\{\begin{array}{ll}
  k<1/2
\left\{\begin{array}{ll}
\int_0^k dp...
\quad
& {\rm with}  \quad p < k  \\
\int_k^{(1-k)}dp...
& {\rm with}\quad p > k \\ 
\end{array} \right.\\
 k>1/2 \quad\int_0^{(1-k)}dp...\quad 
 {\rm with}\quad p < k \\
\end{array} \right. 
\end{displaymath}

\begin{displaymath}
\int_{(1-k)}^1 dp\rightarrow
\left\{\begin{array}{ll}
 k<1/2 \quad\int_{(1-k)}^1dp... \qquad
 {\rm with}\quad p > k \\ 
 k>1/2
\left\{\begin{array}{ll}
\int_{(1-k)}^k dp...
\quad
& {\rm with}  \quad p < k  \\
\int_k^1 dp...
& {\rm with}\quad p > k \\
\end{array} \right.
 \end{array} \right. 
\end{displaymath}

\begin{displaymath}
\int_{(k-1)}^1 dp\rightarrow
\left\{\begin{array}{ll}
 1<k<2 \quad\int_{(k-1)}^1dp... \quad
 {\rm with}\quad p < k \\
\end{array}  \right.
\end{displaymath}
\noindent

Following the scheme (\ref{intscheme}) we can now perform the integration over $p$.
Our exact results are given for particular values of $n_B$. 

\subsection{$n_B=3$}

\begin{displaymath}
|\rho_B (k) |^2_{n_B=3} = 
\frac{A^2 k_D^{9}}{512\pi^4 k_*^{6}}
\left\{\begin{array}{ll}
\frac{4}{9} - \tk + \frac{20\tk^2}{21}- \frac{5\tk^3}{12}+\frac{4\tk^4}{75} + \frac{4\tk^6}{315}-\frac{\tk^9}{1575} 
\quad & {\rm for}  \quad 0 \le \tk \le 1 \\
-\frac{4}{9} - \frac{88}{525\tk} + \frac{13\tk}{15}-\frac{20\tk^2}{21}
+ \frac{17\tk^3}{36} - \frac{4\tk^4}{75} - \frac{4\tk^6}{315}+\frac{\tk^9}{525}
& {\rm for}
\quad 1 \le \tk  \le 2
\end{array} 
\label{scalar_n3}
\right. \,,
\end{displaymath}
\begin{displaymath}
|\Pi_B^{(V)} (k) |^2_{n_B=3} = 
\frac{A^2 k_D^{9}}{256 \pi^4 k_*^{6}}
\left\{\begin{array}{ll}
\frac{28}{135} - \frac{5 \tk}{12} + \frac{296 \tk^2}{735} - \frac{2 \tk^3}{9} + \frac{92 \tk^4}{1575} - \frac{
 32 \tk^6}{10395} + \frac {2 \tk^9}{11025} \quad
& {\rm for}  \quad 0 \le \tk \le 1 \\
-\frac{28}{135} - \frac{32}{24255 \tk^5} + \frac{4}{945 \tk^3} + \frac{44}{525 \tk} + \frac{23 \tk}{60} - \frac{
 296 \tk^2}{735} + \frac{2 \tk^3}{9} - \frac{92 \tk^4}{1575} + \frac{32 \tk^6}{10395} - \frac{
 2 \tk^9}{33075}
& {\rm for}
\quad 1 \le \tk  \le 2
\end{array} 
\label{vector_n3}
\right. \,,
\end{displaymath}
\begin{displaymath}
|\Pi_B^{(T)} (k) |^2_{n_B=3} = 
\frac{A^2 k_D^{9}}{256 \pi^4 k_*^{6}}
\left\{\begin{array}{ll}
\frac{56}{135} - \frac{7 \tk}{6} + \frac{1112 \tk^2}{735} - \frac{127 \tk^3}{144} + \frac{296 \tk^4}{1575} + \frac{
 104 \tk^6}{10395} - \frac{29 \tk^9}{11025} \quad
& {\rm for}  \quad 0 \le \tk \le 1 \\
-\frac{56}{135} +\frac {16}{24255 \tk^5} + \frac{8}{945 \tk^3} + \frac{32}{525 \tk} + \frac{37 \tk}{30} - \frac{
 1112 \tk^2}{735} + \frac{43 \tk^3}{48} - \frac{296 \tk^4}{1575} - \frac{104 \tk^6}{10395} + \frac{
 29 \tk^9} {33075}
& {\rm for}
\quad 1 \le \tk  \le 2
\end{array} 
\label{tensor_n3}
\right. \,.
\end{displaymath}

\subsection{$n_B=2$}
\be
|\rho_B (k) |^2_{n_B=2} = \frac{A^2 k_D^{7}}{512\pi^4 k_*^{4}}
\left[ 
\frac{4}{7} - \tk + \frac{8\tk^2}{15} - \frac{\tk^5}{24} 
+ \frac{11\tk^7}{2240} \right] \nonumber\,,
\label{scalar_n2}
\ee
\be
| \Pi_B^{(V)}(k) |^2_{n_B=2} = \frac{A^2 k_D^{7}}{256 \pi^4 k_*^{4}}
\left[\frac{4}{15} - \frac{5 \tk}{12} + \frac{4 \tk^2}{15} - \frac{\tk^3}{12} + \frac{7 \tk^5}{960} - \frac{\tk^7}{1920} 
 \right] \nonumber\,,
\label{vector_n2}
\ee
\be
| \Pi_B^{(T)}(k) |^2_{n_B=2} = \frac{A^2 k_D^{7}}{256 \pi^4 k_*^{4}}
\left[\frac{8}{15} - \frac{7 \tk}{6} + \frac{16 \tk^2}{15} - \frac{7 \tk^3}{24} - \frac{13 \tk^5}{480} + \frac{
 11 \tk^7}{1920}
 \right] \nonumber\,.
\label{tensor_n2}
\ee

\subsection{$n_B=1$}

\begin{displaymath}
|\rho_B (k) |^2_{n_B=1} = 
\frac{A^2 k_D^{5}}{512\pi^4 k_*^{2}}
\left\{\begin{array}{ll}
\frac{4}{5} - \tk + \frac{\tk^3}{4} - \frac{4}{15} \tk^4 - \frac{\tk^5}{5}  
\quad & {\rm for}  \quad 0 \le \tk \le 1 \\
\frac{8}{15\tk}-\frac{4}{5}+\frac{\tk}{3}+\frac{\tk^3}{4}-\frac{4\tk^4}{15} + \frac{\tk^5}{15}
& {\rm for} \quad 1 \le \tk  \le 2
\end{array} 
\label{scalar_n1}
\right. \,,
\end{displaymath}

\begin{displaymath}
|\Pi_B^{(V)} (k) |^2_{n_B=1} = 
\frac{A^2 k_D^{5}}{256 \pi^4 k_*^{2}}
\left\{\begin{array}{ll}
\frac{28}{75} - \frac{5 \tk}{12} + \frac{4 \tk^2}{35} - \frac{8 \tk^4}{315} + \frac{\tk^5}{50} \quad
& {\rm for}  \quad 0 \le \tk \le 1 \\
- \frac{32}{1575 \tk^5} + \frac{4}{105 \tk^3} + \frac{4}{15 \tk} -\frac{28}{75} + \frac{ \tk}{4} - \frac{
4 \tk^2}{35} + \frac{8 \tk^4}{315} - \frac{ \tk^5}{150} & {\rm for} \quad 1 \le \tk  \le 2
\end{array} 
\label{vector_n1}
\right. ,
\end{displaymath}

\begin{displaymath}
|\Pi_B^{(T)} (k) |^2_{n_B=1} = 
\frac{A^2 k_D^{5}}{256 \pi^4 k_*^{2}}
\left\{\begin{array}{ll}
\frac{56}{75} - \frac{7 \tk}{6} + \frac{64 \tk^2}{105} - \frac{\tk^3}{16} + \frac{8 \tk^4}{63} - \frac{4\tk^5}{25}
 \quad & {\rm for}  \quad 0 \le \tk \le 1 \\
\frac {16}{1575 \tk^5} + \frac{8}{105 \tk^3} -\frac{56}{75}+ \frac{3\tk}{2} - \frac{64 \tk^2}{105} + \frac{\tk^3}{16}-\frac{8\tk^4}{63}-\frac{4 \tk^5}{75}
& {\rm for}\quad 1 \le \tk  \le 2
\end{array} 
\label{tensor_n1}
\right. \,.
\end{displaymath}

\subsection{$n_B=0$}

{\setlength\arraycolsep{2pt}
\begin{eqnarray}
|\rho_B (k) |^2_{n_B=0} &=&\frac{A^2 k_D^{3}}{512\pi^4}  \bigg[\frac{29}{24}-\frac{17\tk}{16}-\frac{7\tk^2}{8}+\frac{53\tk^3}{96}+
\frac{\pi^2\tk^3}{24}-\frac{\log |1-\tk|}{8\tk}+\frac{\tk\log |1-\tk|}{2}-\frac{3\tk^3\log |1-\tk|}{8}\nonumber\\
&&+\frac{\tk^3 \log |1-\tk|\log \tk}{2}-\frac{\tk^3 \log^2 \tk}{4}-\frac{\tk^3{\rm PolyLog}[2,\frac{-1+\tk}{\tk}]}{2} \bigg] \nonumber\,,
\end{eqnarray}}

{\setlength\arraycolsep{2pt}
\begin{eqnarray}
|\Pi_B^{(V)} (k) |^2_{n_B=0} &=&\frac{A^2 k_D^{3}}{256 \pi^4}  \bigg[\frac{53}{96}+\frac{1}{32\tk^4}+\frac{1}{64\tk}-\frac{1}{32\tk^2}-\frac{5}{384\tk}-\frac{29\tk}{64}-\frac{5\tk^2}{96}+\frac{55\tk^3}{768}+\frac{\log |1-\tk|}{32\tk^5}-\frac{\log |1-\tk|}{24\tk^3}\nonumber \\
&&{}-\frac{\log |1-\tk|}{16\tk}+\frac{\tk \log |1-\tk|}{8}-\frac{5\tk \log |1-\tk|}{96} \bigg] \nonumber\,,
\end{eqnarray}}

{\setlength\arraycolsep{2pt}
\begin{eqnarray}
|\Pi_B^{(T)} (k) |^2_{n_B=0} &=&\frac{A^2 k_D^{3}}{256 \pi^4}  \bigg[\frac{293}{192}-\frac{1}{64\tk^4}-\frac{1}{128\tk}-\frac{17}{192\tk^2}-\frac{35}{768\tk}-\frac{397\tk}{384}-\frac{17\tk^2}{192}+\frac{181\tk^3}{1536}+\frac{\pi^2\tk^3}{96}-\frac{\log |1-\tk |}{64\tk^5}-\frac{\log |1-\tk|}{12\tk^3}\nonumber \\
&&{}+\frac{5\log |1-\tk|}{16\tk}-\frac{\tk \log |1-\tk|}{4}+\frac{7\tk^3 \log |1-\tk|}{192}+\frac{\tk^3\log |1-\tk |\log \tk}{8}-\frac{\tk^3\log^2 \tk}{16} - \frac{\tk^3 {\rm PolyLog} [2\frac{-1+\tk}{\tk}]}{8}\bigg] \nonumber\,.
\end{eqnarray}}

\subsection{$n_B=-1$}

\begin{displaymath}
|\rho_B (k) |^2_{n_B=-1} = \frac{A^2 k_D k_*^2}{512\pi^4} 
\left\{\begin{array}{ll}
4 - 5 \tk + \frac{4 \tk^2}{3} +\frac{\tk^3}{4}
\quad 
& {\rm for}  \quad 0 \le \tk \le 1 \\ 
-4 + \frac{8}{3\tk} +3\tk - \frac{4 \tk^2}{3} +\frac{\tk^3}{4}
& {\rm for}  
\quad 1 \le \tk  \le 2
\end{array} \right. \,,
\label{scalarn-1}
\end{displaymath}

\begin{displaymath}
|\Pi_B^{(V)} (k) |^2_{n_B=-1} =\frac{A^2 k_D k_*^2}{256 \pi^4}
\left\{\begin{array}{ll}
\frac{28}{15} - \frac{7\tk}{4} + \frac{16\tk^2}{105}
\quad 
& {\rm for}  \quad 0 \le \tk \le 1 \\ 
-\frac{28}{15} + \frac{32}{105 \tk^5} - \frac{4}{15 \tk^3} + \frac{4}{3 \tk} + \frac{11 \tk}{12} - \frac{
 16 \tk^2}{105}
& {\rm for}  
\quad 1 \le \tk  \le 2
\end{array} \right. \,,
\label{vectorn-1}
\end{displaymath}
\begin{displaymath}
|\Pi_B^{(T)} (k) |^2_{n_B=-1} =\frac{A^2 k_D k_*^2}{256 \pi^4}
\left\{\begin{array}{ll}
\frac{56}{15} - \frac{5 \tk}{2} - \frac{8 \tk^2}{105} + \frac{\tk^3}{16}
\quad 
& {\rm for}  \quad 0 \le \tk \le 1 \\ 
-\frac{56}{15} - \frac{16}{105 \tk^5} - \frac{8}{15 \tk^3} + \frac{16}{3 \tk} + \frac{\tk}{6} + \frac{
 8 \tk^2}{105} + \frac{\tk^3}{16}
& {\rm for}  
\quad 1 \le \tk  \le 2
\end{array} \right. \,.
\label{tensorn-1}
\end{displaymath}

\subsection{$n_B=-3/2$}
\begin{displaymath}
|\rho_B (k) |^2_{n_B=-3/2} = \frac{A^2 k_*^3}{512\pi^4} 
\left\{\begin{array}{ll}
\frac{232}{45\sqrt{1-\tk}}+\frac{88}{15\tk}-\frac{88}{15\sqrt{1-\tk}\tk}-2\pi +\frac{4\tk}{3}
-\frac{32\tk}{45\sqrt{1-\tk}}+\frac{64\tk^2}{45\sqrt{1-\tk}} \\
+\frac{\tk^3}{9} +8\log[1+\sqrt{1-\tk}]-4\log \tk
\quad & {\rm for}  \quad 0 \le \tk \le 1 \\ 
-\frac{232}{45\sqrt{-1+\tk}}+\frac{88}{15\tk}+\frac{88}{15\sqrt{-1+\tk}\tk} +\frac{4\tk}{3}
+\frac{32\tk}{45\sqrt{-1+\tk}}-\frac{64\tk^2}{45\sqrt{-1+\tk}} \\
+\frac{\tk^3}{9} -4\arctan \big[\frac{1}{\sqrt{-1+\tk}}\big]+4\arctan \big[\sqrt{-1+\tk}\big]
& {\rm for}  
\quad 1 \le \tk  \le 2
\end{array} \right. \,,
\label{n-1}
\end{displaymath}

\begin{displaymath}
|\Pi_B^{(V)} (k) |^2_{n_B=-3/2} = \frac{A^2 k_*^3}{256 \pi^4} 
\left\{\begin{array}{ll}
\frac{4936}{1755\sqrt{1-\tk}}+\frac{1024}{2925\tk^5}-\frac{1024}{2925\sqrt{1-\tk}\tk^5}-\frac{14\pi}{15} +\frac{512}{2925\sqrt{1-\tk}\tk^4}\\ -\frac{32}{135\tk^3}+\frac{2464}{8775\sqrt{1-\tk}\tk^3}-\frac{848}{8775\sqrt{1-\tk}\tk^2} +\frac{44}{15\tk}
-\frac{5176}{1755\sqrt{1-\tk}\tk}+\frac{\tk}{3} \\
-\frac{224\tk}{1755\sqrt{1-\tk}}+\frac{448\tk^2}{1755\sqrt{1-\tk}}+\frac{56\log [1+\sqrt{1-\tk}]}{15}-\frac{28\log \tk}{15}
\quad & {\rm for}  \quad 0 \le \tk \le 1 \\ 
-\frac{4936}{1755\sqrt{-1+\tk}}+\frac{1024}{2925\tk^5}+\frac{1024}{2925\sqrt{-1+\tk}\tk^5} -\frac{512}{2925\sqrt{-1+\tk}\tk^4}\\ 
-\frac{32}{135\tk^3}-\frac{2464}{8775\sqrt{-1+\tk}\tk^3}+\frac{848}{8775\sqrt{-1+\tk}\tk^2} +\frac{44}{15\tk}
+\frac{5176}{1755\sqrt{-1+\tk}\tk}+\frac{\tk}{3} \\
+\frac{224\tk}{1755\sqrt{-1+\tk}}-\frac{448\tk^2}{1755\sqrt{-1+\tk}}-\frac{28\arctan \big[\frac{1}{\sqrt{-1+\tk}}\big]}{15}+\frac{28\arctan \big[\sqrt{-1+\tk}\big]}{15}
& {\rm for}  
\quad 1 \le \tk  \le 2
\end{array} \right. \,,
\end{displaymath}

\begin{displaymath}
|\Pi_B^{(T)} (k) |^2_{n_B=-3/2} = \frac{A^2 k_*^3}{256 \pi^4} 
\left\{\begin{array}{ll}
\frac{16304}{1755\sqrt{1-\tk}}-\frac{512}{2925\tk^5}+\frac{512}{2925\sqrt{1-\tk}\tk^5}-\frac{28\pi}{15} -\frac{256}{2925\sqrt{1-\tk}\tk^4}\\ -\frac{64}{135\tk^3}+\frac{3968}{8775\sqrt{1-\tk}\tk^3}-\frac{2176}{8775\sqrt{1-\tk}\tk^2} +\frac{28}{3\tk}
-\frac{16496}{1755\sqrt{1-\tk}\tk}-\frac{2\tk}{3} \\
+\frac{64\tk}{351\sqrt{1-\tk}}-\frac{128\tk^2}{351\sqrt{1-\tk}}+\frac{\tk^3}{36}+\frac{112\log [1+\sqrt{1-\tk}]}{15}-\frac{56\log \tk}{15}
\quad & {\rm for}  \quad 0 \le \tk \le 1 \\ 
-\frac{16304}{1755\sqrt{-1+\tk}}-\frac{512}{2925\tk^5}-\frac{512}{2925\sqrt{-1+\tk}\tk^5} +\frac{256}{2925\sqrt{-1+\tk}\tk^4}\\ 
-\frac{64}{135\tk^3}-\frac{3968}{8775\sqrt{-1+\tk}\tk^3}+\frac{2176}{8775\sqrt{-1+\tk}\tk^2} +\frac{28}{3\tk}
+\frac{16496}{1755\sqrt{-1+\tk}\tk}-\frac{2\tk}{3} \\
-\frac{64\tk}{351\sqrt{-1+\tk}}+\frac{128\tk^2}{351\sqrt{-1+\tk}}+\frac{\tk^3}{36}-\frac{56\arctan \big[\frac{1}{\sqrt{-1+\tk}}\big]}{15}+\frac{56\arctan \big[\sqrt{-1+\tk}\big]}{15}
& {\rm for}  
\quad 1 \le \tk  \le 2
\end{array} \right. \,.
\end{displaymath}

\subsection{$n_B=-5/2$}

\begin{displaymath}
|\rho_B (k) |^2_{n_B=-5/2} = \frac{A^2 k_*^5}{512\pi^4 k_D^2} 
\bigg[-\frac{32}{75\sqrt{|1-\tk|}}+\frac{272}{25\sqrt{|1-\tk |}\tk^2}+\frac{88}{15\tk}-\frac{848}{75\sqrt{|1-\tk |}\tk}-\frac{4\tk}{5}
+\frac{64\tk}{75\sqrt{|1-\tk |}} +\frac{\tk^3}{25}\bigg]
\end{displaymath}

{\setlength\arraycolsep{2pt}
\begin{eqnarray}
|\Pi_B^{(V)} (k) |^2_{n_B=-5/2} &=& \frac{A^2 k_*^5}{256 \pi^4 k_D^2} 
\bigg[-\frac{32}{231\sqrt{|1-\tk|}}-\frac{1024}{1155\tk^5}+\frac{1024}{1155\sqrt{|1-\tk |}\tk^5}-\frac{512}{1155\sqrt{|1-\tk |}\tk^4}+\frac{32}{105\tk^3}\nonumber\\
&&-\frac{32}{77\sqrt{|1-\tk |}\tk^3}
+\frac{896}{165\sqrt{|1-\tk |}\tk^2}+\frac{44}{15\tk}-\frac{6464}{1155\sqrt{|1-\tk |}\tk}
-\frac{\tk}{5}+\frac{64\tk}{231\sqrt{|1-\tk |}}\bigg]\nonumber
\end{eqnarray}}

{\setlength\arraycolsep{2pt}
\begin{eqnarray}
|\Pi_B^{(T)} (k) |^2_{n_B=-5/2} &=& \frac{A^2 k_*^5}{256 \pi^4 k_D^2} 
\bigg[\frac{1984}{5775\sqrt{|1-\tk |}}+\frac{512}{1155\tk^5}-\frac{512}{1155\sqrt{|1-\tk |}\tk^5}+\frac{256}{1155\sqrt{|1-\tk |}\tk^4}+\frac{64}{105\tk^3}\nonumber\\
&&-\frac{128}{231\sqrt{|1-\tk |}\tk^3}
+\frac{117728}{5775\sqrt{|1-\tk |}\tk^2}+\frac{28}{3\tk}-\frac{37088}{1925\sqrt{|1-\tk |}\tk}
+\frac{2\tk}{5}-\frac{3968\tk}{5775\sqrt{|1-\tk |}}+\frac{\tk^3}{100}\bigg]\nonumber
\end{eqnarray}}

\setcounter{equation}{0}
%\end{widetext}
\section{Scalar part of the Lorentz Force}

In order to compute the scalar contribution 
of a SB of PMFs to the cosmological perturbations, the 
convolution for the scalar part of the Lorentz Force power spectrum 
is also necessary.
The scalar anisotropic stress can be obtained directly from 
its relation with the Lorentz force and the magnetic energy density in Eq.(\ref{stress_B})
.
We report here the result for the Lorentz force convolution:
\be
|L (k)|^2 =\frac{1}{1024 \pi^5 a^8}\int &d^3 p& P_B(p) \, P_B(|{\mathbf{k}}
-{\mathbf{p}}|) [1 + \mu^2 + 4 \gamma \beta(\gamma \beta - \mu)] \,,
\label{spectrum_LF}
\ee
with the magnetic 
field power spectrum in Eq. (\ref{PSpectrum}) for particular values of $n_B$.

\subsection{$n_B=3$}

\begin{displaymath}
|L (k) |^2_{n_B=3} = \frac{A^2 k_D^9}{512\pi^4 k_*^6}
\left\{\begin{array}{ll}
\frac{44}{135} - \frac{2 \tk}{3} + \frac{556 \tk^2}{735} - \frac{4 \tk^3}{9} + \frac{164 \tk^4}{1575} +\frac{4 \tk^6}{2079} -\frac{11 \tk^9}{11025}
\quad
& {\rm for}  \quad 0 \le \tk \le 1 \\
-\frac{44}{135} +\frac{ 64}{24255 \tk^5} - \frac{16}{945 \tk^3} +\frac{ 88}{525 \tk} + \frac{2 \tk}{3} -\nonumber\\
\frac{ 556 \tk^2}{735} + \frac{4 \tk^3}{9} -\frac{164 \tk^4}{1575} - \frac{4 \tk^6}{2079} + \frac{11 \tk^9}{33075} 
& {\rm for}
\quad 1 \le \tk  \le 2
\end{array} \right. .
\end{displaymath}

\subsection{$n_B=2$}

\begin{displaymath}
|L (k) |^2_{n_B=2} = \frac{A^2 k_D^7}{512\pi^4 k_*^4}
\left[\frac{44}{105} - \frac{2 \tk}{3} + \frac{8 \tk^2}{15} -\frac{\tk^3}{6} - \frac{\tk^5}{240} + \frac{13 \tk^7}{6720}
 \right] \,.
\end{displaymath}

\subsection{$n_B=1$}

\begin{displaymath}
|L (k) |^2_{n=1} = \frac{A^2 k_D^5}{512\pi^4 k_*^2}
\left\{\begin{array}{ll}
\frac{44}{75} - \frac{2 \tk}{3} + \frac{32 \tk^2}{105} +\frac{4 \tk^4}{315} - \frac{k^5}{25}
\quad
& {\rm for}  \quad 0 \le \tk \le 1 \\
  -\frac{44}{75} + \frac{64}{1575 \tk^5} - \frac{16}{105\tk^3} +\frac{8}{
  15 \tk} +\frac{2 \tk}{3} - \frac{32 \tk^2}{105} - \frac{4 \tk^4}{315} + \frac{\tk^5}{75}
& {\rm for}
\quad 1 \le \tk  \le 2
\end{array} \right. \,.
\end{displaymath}

\subsection{$n_B=0$}
{\setlength\arraycolsep{2pt}
\begin{eqnarray}
|L (k) |^2_{n_B=0} &=& \frac{A^2 k_D^3}{512\pi^4}
\bigg[\frac{43}{48} -\frac{ 1}{16 \tk^4} - \frac{1}{32 \tk^3} + \frac{7}{48 \tk^2} + \frac{13}{192 \tk} - \frac{67 \tk}{
   96} +\frac{ \tk^2}{48} +\frac{17 \tk^3}{384} -\frac{ \log |1 - k|}{16 \tk^5} \nonumber\\
&&+ \frac{\log |1 - k|}{6 \tk^3} -\frac{ \log |1 - k|}{8 \tk} +\frac{\tk^3 \log |1 - k|}{48}\bigg] \nonumber\,.
\end{eqnarray}}

\subsection{$n_B=-1$}

\begin{displaymath}
|L (k) |^2_{n_B=-1} = \frac{A^2 k_D k_*^2}{512\pi^4}
\left\{\begin{array}{ll}
 \frac{44}{15} - 2 \tk - \frac{4 \tk^2}{105}
\quad
& {\rm for}  \quad 0 \le \tk \le 1 \\
-\frac{44}{15} -\frac{ 64}{105 \tk^5} +\frac{16}{15 \tk^3} + \frac{8}{3 \tk} +\frac{
   2 \tk}{3} +\frac{4 \tk^2}{105}
& {\rm for}
\quad 1 \le \tk  \le 2
\end{array} \right. \,.
\end{displaymath}

\subsection{$n_B=-3/2$}

\begin{displaymath}
|L (k) |^2_{n=-3/2} = \frac{A^2 k_*^3}{512\pi^4} 
\left\{\begin{array}{ll}
\frac{10616}{1755\sqrt{1-\tk}}-\frac{2048}{2925\tk^5}+\frac{2048}{2925\sqrt{1-\tk}\tk^5}-\frac{22\pi}{15}+\frac{128}{135\tk^3} -\frac{9088}{8775\sqrt{1-\tk}\tk^3}\\ +\frac{88}{15\tk}-\frac{10136}{1775\sqrt{1-\tk}\tk} +\frac{32\tk}{1755\sqrt{1-\tk}}-\frac{64\tk^2}{1775\sqrt{1-\tk}}+\frac{88\log [1+\sqrt{1-\tk}]}{15}-\frac{44\log \tk}{15}
\quad & {\rm for}  \quad 0 \le \tk \le 1 \\ 
-\frac{10616}{1755\sqrt{-1+\tk}}-\frac{2048}{2925\tk^5}-\frac{2048}{2925\sqrt{-1+\tk}\tk^5} +\frac{1024}{2925\sqrt{-1+\tk}\tk^4}\\ 
+\frac{128}{135\tk^3}+\frac{9088}{8775\sqrt{-1+\tk}\tk^3}-\frac{3776}{8775\sqrt{-1+\tk}\tk^2} +\frac{88}{15\tk}
+\frac{10136}{1755\sqrt{-1+\tk}\tk} \\
-\frac{32\tk}{1775\sqrt{-1+\tk}}+\frac{64\tk^2}{1775\sqrt{-1+\tk}}-\frac{44\arctan \big[\frac{1}{\sqrt{-1+\tk}}\big]}{15}+\frac{44\arctan \big[\sqrt{-1+\tk}\big]}{15}
& {\rm for}  
\quad 1 \le \tk  \le 2
\end{array} \right. \,.
\end{displaymath}

\subsection{$n_B=-5/2$}

{\setlength\arraycolsep{2pt}
\begin{eqnarray}
|L (k) |^2_{n_B=-5/2} &=& \frac{A^2 k_*^5}{512\pi^4 k_D^2} 
\bigg[-\frac{32}{1155\sqrt{|1-\tk|}}+\frac{2048}{1155\tk^5}-\frac{2048}{1155\sqrt{|1-\tk |}\tk^5}+\frac{1024}{1155\sqrt{|1-\tk |}\tk^4}-
\frac{128}{105\tk^3}\nonumber\\
&&+\frac{1664}{1155\sqrt{|1-\tk |}\tk^3}
+\frac{12976}{1155\sqrt{|1-\tk |}\tk^2}+\frac{88}{15\tk}-\frac{13648}{1155\sqrt{|1-\tk |}\tk}
+\frac{64\tk}{1155\sqrt{|1-\tk |}}\bigg]\nonumber
\end{eqnarray}}
\end{onecolumn}
%\end{widetext}
%\bigskip
%\bigskip
%\noindent

{\bf Acknowledgements.} 
We wish to thank Chiara Caprini, Ruth Durrer, Antony Lewis 
and Kandaswamy Subramanian for comments and discussions.
This work has been done in the framework of the Planck LFI activities and 
is partially supported by 
ASI contract Planck LFI Activity of Phase E2.

\end{document}